\documentclass[twocolumn]{aastex631}
\usepackage{threeparttable}
\usepackage{amsmath}

\newcommand{\um}{$\mu$m}
\newcommand{\ai}{$\mbox{\normalfont\AA}$ }

\newcommand\dsmg{$\delta_{\rm{SMG}}$}
\newcommand{\cii}{[C\thinspace{\sc ii}]}

\received{XXX}
\revised{XXX}
\accepted{XXX}

\submitjournal{ApJ}

\shorttitle{SHERRY - II: the environment of $z\sim6$ quasars at sub-millimeter band}
\shortauthors{Li et al.}
\graphicspath{{./}{figures/}}

\begin{document}

\title{(SHERRY) JCMT-SCUBA2 High rEdshift bRight quasaR surveY - II: the environment of $z\sim6$ quasars at sub-millimeter band}

\correspondingauthor{Ran Wang}
\email{rwangkiaa@pku.edu.cn}

\author[0000-0002-3119-9003]{Qiong Li}
\affil{Department of Astronomy, School of Physics, Peking University, Beijing 100871, P. R. China}
\affil{Kavli Institute for Astronomy and Astrophysics, Peking University, Beijing, 100871, P. R. China}
\affil{Jodrell Bank Centre for Astrophysics, University of Manchester, Oxford Road, Manchester M13 9PL, UK}
\author[0000-0003-4956-5742]{Ran Wang}
\affil{Department of Astronomy, School of Physics, Peking University, Beijing 100871, P. R. China}
\affil{Kavli Institute for Astronomy and Astrophysics, Peking University, Beijing, 100871, P. R. China}
\author[0000-0003-3310-0131]{Xiaohui Fan}
\affil{University of Arizona (Steward Observatory), United States}
\author[0000-0002-7350-6913]{Xue-Bing Wu}
\affil{Department of Astronomy, School of Physics, Peking University, Beijing 100871, P. R. China}
\affil{Kavli Institute for Astronomy and Astrophysics, Peking University, Beijing, 100871, P. R. China}
\author[0000-0003-4176-6486]{Linhua Jiang}
\affil{Kavli Institute for Astronomy and Astrophysics, Peking University, Beijing, 100871, P. R. China}
\author[0000-0002-2931-7824]{Eduardo Ba\~nados}
\affil{Max-Planck-Institut f\"{u}r Astronomie, K\"{o}nigstuhl 17, D-69117, Heidelberg, Germany}
\author[0000-0001-9024-8322]{Bram Venemans}
\affil{Max-Planck-Institut f\"{u}r Astronomie, K\"{o}nigstuhl 17, D-69117, Heidelberg, Germany}
\author[0000-0002-1478-2598]{Yali Shao}
\affil{Department of Astronomy, School of Physics, Peking University, Beijing 100871, P. R. China}
\affil{Kavli Institute for Astronomy and Astrophysics, Peking University, Beijing, 100871, P. R. China}
\author[0000-0002-1815-4839]{Jianan Li}
\affil{Department of Astronomy, School of Physics, Peking University, Beijing 100871, P. R. China}
\affil{Kavli Institute for Astronomy and Astrophysics, Peking University, Beijing, 100871, P. R. China}
\author{Jeff Wagg}
\affil{Square Kilometre Array Observatory, Lower Withington, Macclesfield, Cheshire SK11 9FT, UK}
\author[0000-0002-2662-8803]{Roberto Decarli}
\affil{INAF -- Osservatorio di Astrofisica e Scienza dello Spazio di Bologna, via Gobetti 93/3, I-40129, Bologna, Italy.}
\author[0000-0002-5941-5214]{Chiara Mazzucchelli}
\affil{European Southern Observatory, Alonso de Cordova 3107, Vitacura, Region Metropolitana, Chile}
\author[0000-0002-4721-3922]{Alain Omont}
\affil{Sorbonne Universit\'e, UPMC Universit\'e Paris 6 and CNRS, UMR 7095, Institut d'Astrophysique de Paris, France}
\author[0000-0002-1707-1775]{Frank Bertoldi}
\affil{Universit\"{a}t Bonn (Argelander-Institut f\"{u}r Astronomie), Germany}
\author[0000-0001-9487-8583]{Sean Johnson}
\affil{Department of Astronomy, University of Michigan, 311 West Hall, 1085 S. University Ave, Ann Arbor, MI, 48109-1107, U.S.A.}
\author[0000-0003-1949-7638]{Christopher J. Conselice}
\affiliation{Jodrell Bank Centre for Astrophysics, University of Manchester, Oxford Road, Manchester M13 9PL, UK}
\author[0000-0001-6469-1582]{Chengpeng Zhang}
\affiliation{Department of Physics and Astronomy, Texas A\&M University, College Station, TX 77843-4242, USA}
\affiliation{George P. and Cynthia Woods Mitchell Institute for Fundamental Physics and Astronomy, Texas A\&M University, College Station, TX 77843-4242, USA}


\begin{abstract}

The formation of the first supermassive black holes is expected to have occurred in some most pronounced matter and galaxy overdensities in the early universe. We have conducted a sub-mm wavelength continuum survey of 54 $z\sim6$ quasars using the Submillimeter Common-User Bolometre Array-2 (SCUBA2) on the James Clerk Maxwell Telescope (JCMT)
to study the environments around $z \sim 6$ quasars. We identified 170 submillimeter galaxies (SMGs) with above 3.5$\sigma$ detections at 450 or 850 \um\, maps.
Their FIR luminosities are 2.2 - 6.4 $\times$ 10$^{12} L_{\odot}$, and star formation rates are $\sim$ 400 - 1200 M$_{\odot}$ yr$^{-1}$. We also calculated the SMGs differential and cumulative number counts in a combined area of $\sim$ 620 arcmin$^2$. To a $4\sigma$ detection (at $\sim$ 5.5 mJy), SMGs overdensity is $0.68^{+0.21}_{-0.19}$($\pm0.19$), exceeding the blank field source counts by a factor of 1.68. We find that 13/54 quasars show overdensities (at $\sim$ 5.5 mJy) of $\delta_{SMG}\sim$ 1.5 - 5.4. The combined area of these 13 quasars exceeds the blank field counts with the overdensity to 5.5 mJy of \dsmg $\sim$ $2.46^{+0.64}_{-0.55}$($\pm0.25$) in the regions of $\sim$ 150 arcmin$^2$. However, the excess is insignificant on the bright end (e.g., 7.5 mJy). We also compare results with previous environmental studies of Lyman alpha emitters (LAEs) and Lyman-Break Galaxies (LBGs) on a similar scale. Our survey presents the first systematic study of the environment of quasars at $z\sim6$. The newly discovered SMGs provide essential candidates for follow-up spectroscopic observations to test whether they reside in the same large-scale structures as the quasars and search for protoclusters at an early epoch.

\end{abstract}



\section{Introduction} \label{sec:intro}

More than 200 quasars at $z >$ 5.6 were discovered in optical and near-infrared surveys, such as SDSS \citep[e.g.][]{Jiang2008,Fan2006}, CFHQS \citep[e.g.][]{Willott2007,Willott2010}, UKIDSS \citep[e.g.][]{Venemans2007,Mortlock2011}, VIKING \citep[e.g.][]{Venemans2013,Venemans2015}, VST-ATLAS \citep{Carnall2015}, DES \citep{Reed2015,Yang2019}, HSC \citep{Matsuoka2016}, Pan-STARRS1 \citep[e.g. PS1;][]{Banados2014,Banados2016,Wang2019}, etc (e.g. \citealt{Yang2021}).
Galaxy formation simulations and cosmological models predict that high redshift quasars reside in the most massive dark matter halos and grow quickly in the galaxy overdensity environments formed by the hierarchical merging of galaxies \citep[e.g.][]{RomanoDiaz2011,Costa2014}.
Discovering the overdensities of galaxies associated with $z\sim6$ quasars will trace mass assembly and probe the evolution of protoclusters and large-scale structures in the early universe.
Hence, it is a chance to investigate the environment where QSOs hosting SMBHs reside and to search for galaxy protoclusters close to the end of cosmic reionization.

Over the last twenty years, sub-mm surveys around AGNs and galaxies have advanced our understanding of the galaxy environment at the cosmic time.
The first one was uncovered by SCUBA (Submillimeter Common-User Bolometre Array) surveys on JCMT (the James Clerk Maxwell Telescope, \citealt{Smail1997}). Subsequently, a series of papers have shown evidence for over-densities of submillimeter galaxies (SMGs) at $z=2\sim3$ in the vicinities of luminous radio galaxies (e.g., \citealt{DeBreuck2004,Greve2007}) and dust absorbed QSOs \citep{Stevens2004, Priddey2007}.
The clustering of SMGs at $z\sim2$ around luminous QSOs or galaxies indicates evolutionary scenarios in which powerful starbursts and QSOs occur in the same systems \citep{Hickox2012}.
And \citet{Priddey2008} also reported deep sub-mm images with a large field of view for three $z>5$ quasars in SCUBA observations.
However, the observational evidence of large-scale structure close to $z \sim 6$ quasars is elusive, limited by a relatively small quasar sample size and strong selection biases of quasars.

There are many different approaches for tracing overdense regions around quasars or galaxies at $z\sim6$, such as Lyman-Break Galaxies (LBGs) selected via broad-band optical/NIR colors \citep{Stiavelli2005,Morselli2014} or Lyman alpha emitters (LAEs) selected via narrow/broadband observations \citep[e.g.,][]{Mazzucchelli2017b,Ota2018}.
Despite the huge investment of optical or near-infrared (NIR) telescope time, the galaxy environments surrounding $z\sim6$ quasars remain unconstrained.
Overdensities of galaxies were discovered with deep optical imaging around several quasars at $4<z<6$ (e.g., the $z=6.28$ quasar SDSS J1030+0524 in \citealt{Stiavelli2005}, and the $z=5.81$ radio-loud quasar SDSS J0836+0054 in \citealt{ZhengW2006} and \citealt{Ajiki2006}).
However, some works failed to find any significant galaxy overdensity (e.g., LAEs surrounding PSO J215-16 at $z=5.73$ in \citealt{Mazzucchelli2017b}, and CFHQS J2329-0301 in \citealt{Goto2017}).
Integral field spectroscopic investigations at radio/sub-mm wavelengths in quasar fields both with ALMA \citep{Decarli2017,Willott2017,Venemans2019} and NOEMA \citep{Wang2011a,Omont2013} revealed the presence of close ($<$ 50\,kpc, $<$ 1cMpc) companion galaxies in the immediate proximity of the first massive black holes. It only traces the small-scale clustering at close separation from the central quasars.

To test whether the companion galaxies identified around some $z \sim 6$ quasars are a small-scale phenomenon or the expression of the long-sought large-scale overdensities postulated by the models, in 2016-2018 we mapped the fields around 54 quasars at $z \sim 6$ over a much larger field-of-view ($\sim$15\arcmin\, in diameter) using SCUBA-2 (Submillimeter Common-User Bolometre Array-2) on JCMT, named as `JCMT-SCUBA2 High rEdshift bRight quasaR surveY' (`SHERRY' survey).
SCUBA-2 has a field of view of 64 sq-arcminutes (only 4.3 sq-arcmins for SCUBA) - a factor of 16 times improvement over SCUBA \citep{Chapin2013}. And it has the fastest mapping speed at 450 \um\, and 850 \um, which makes searching for SMGs more efficient \citep{Chapin2013}. The SHERRY survey is the first to build a large, statistical sample of $z\sim6$ quasars for sub-mm observation, including $\sim20$\% of the $z\sim6$ quasars discovered to date.

In the first paper of this survey (SHERRY-I, \citealt{Qiong2020}), we reported that about 30\% (16/54) quasars were detected with a typical 850$\mu$m RMS sensitivity of 1.2 $\rm mJy\,beam^{-1}$ ($>3.5\sigma$).
The broadband SED for $z\sim6$ quasars is similar to that at $z\sim2$, which indicates there is probably no evolution of quasars' properties with redshift. The optical/near-infrared\,(NIR) spectra of these objects show 11\%\,(6/54) of the sources have weak ultraviolet\,(UV) emission line features. The SHERRY survey also indicates that quasars with sub-mm detections tend to have weaker emission lines than quasars with non-detections. The weak-line quasar at $z \sim 6$ may be a young AGN-galaxy system \citep{Qiong2020}, in which the broad line region is starting to develop slowly and have some unusual properties.



In this paper, we present JCMT/SCUBA-2 observations of the environment around the quasars at $5.6 < z < 7.1$ and search for galaxy protoclusters close to the end of cosmic reionization.
We describe the SCUBA2 observations and data reduction in Section \ref{section 2}.
In Section \ref{section 3}, we introduce our method to extract SMGs candidates and calculate the number counts.
We present the number counts results and the discussions in Section \ref{section 4}.
Finally, we give a summary of the main results in Section \ref{section 5}.
We adopt a $\Lambda$-model cosmology with $H_0 = 71 $ km s$^{-1}$ Mpc$^{-1}$, $\Omega_M = 0.27$ and $\Omega_{\Lambda} = 0.73$ \citep{Spergel2007} throughout this paper.

\begin{figure}
\centering
\includegraphics[width = \linewidth]{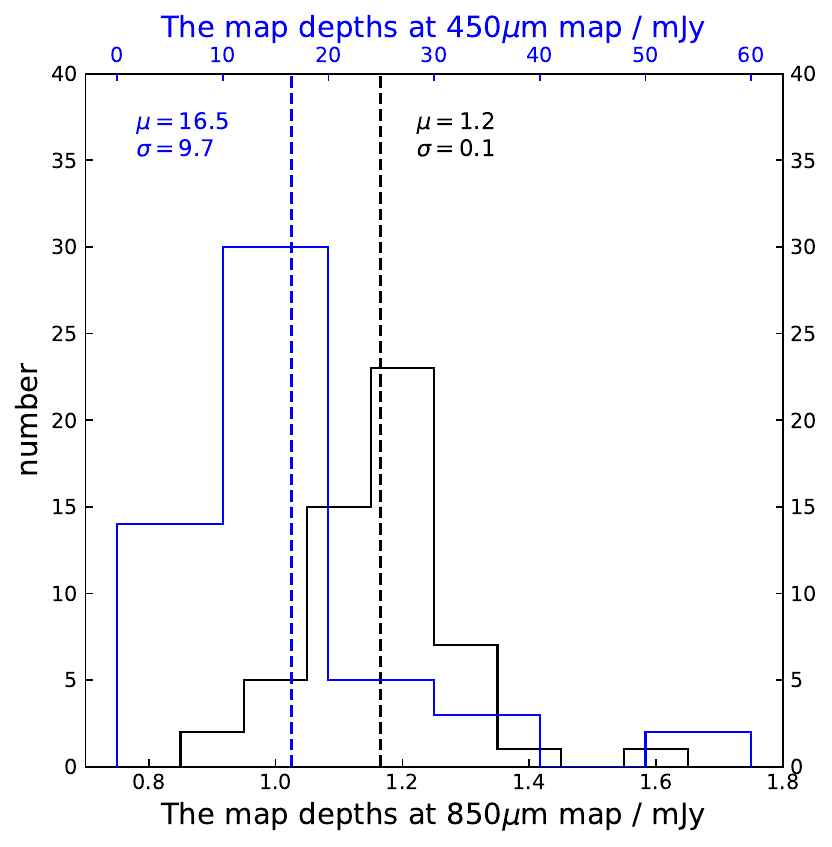}
\caption{ The distrubution of the map depth at 850 and 450 \um.}
\label{figure_8}
\end{figure}

\section{SCUBA2 observations and data reduction}\label{section 2}
We observed 54 quasars at $5.7 < z < 7.1$, which have rest-frame 1450 \ai absolute magnitudes of $M_{\rm 1450}\le-25.4$.
The observations were carried out with the SCUBA2 camera on the JCMT in Hawaii from August 2015 to January 2018 (Program ID: M15BI055, M16AP013, M17AP062, M17BP034).
We used `the constant velocity Daisy scan' observation mode (`CV DAISY' mode) with the field of view of 15\arcmin\, in diameter.
This mode is for small and compact sources of order 3-arcmin or less.
The most uniform and best rms region is the central one with a diameter of about 4\arcmin\,, corresponding to comoving 9.8 h$^{-1}$Mpc comoving. This covers typical protocluster scales at $z\sim6$ (e.g., \citealt{Overzier2009}).

SCUBA2 observes two bands, 450 \um\, and 850 \um\, simultaneously.
The main beam sizes of SCUBA-2 are 7.9\arcsec\, at 450 \um\, and 14.9\arcsec\, at 850 \um; with the spatial resolution of 4\arcsec\,/ pixel and 2\arcsec\,/ pixel, respectively.
The observations were carried out in grade 2 / grade 3 weather conditions with precipitable water vapor (PWV) of 0.83 - 2.58 mm at zenith (atmospheric optical depth $0.05\rm<\rm \tau_{225GHz}\rm<0.12$).
We observed each target in four to six $\sim$ 30-minute scans with a total on-source time of 2$-$3 hours to reach the sensitivity of 1.2\,mJy at 850\um\,. We present the details of the observations in \citet{Qiong2020}.
The calibrated sources were observed before and after the target sources and selected from the James Clerk Maxwell Telescope (JCMT) calibrator list \citep{Dempsey2013}. Calibration sources included Mars, Uranus, Neptune, etc.

We reduced the data using the STARLINK SCUBA-2 science pipeline with the configuration file of `dimmconfig blank field.lis', which is the recipe for processing maps containing the faint compact sources \citep{Chapin2013}.
We processed each complete observation separately to produce an image and calibrated it with the flux conversion factors (FCFs) to mJy/beam.
We adopted the default FCFs value of 537$\pm$24 Jy pW$^{-1}$ beam$^{-1}$ for 850 \um\, map and 491$\pm$67 Jy pW$^{-1}$ beam$^{-1}$ for 450 \um\, map \citep{Dempsey2013}.
We combined all images for a given source into a single image using inverse-variance weighting.
Using this recipe, we further processed the output map with a beam-match filtered of a 15\arcsec\, FWHM Gaussian at 850 \um\, band.
The S/N is taken to enhance point sources, which is suitable for searching for SMGs at $z \sim 6$.
The final 850- and 450-\um\, maps have typical noise levels of 1.2 and 16.5 $\rm mJy\,beam^{-1}$, shown in Figure~\ref{figure_8}.

\begin{figure}
\centering
\includegraphics[width = \linewidth]{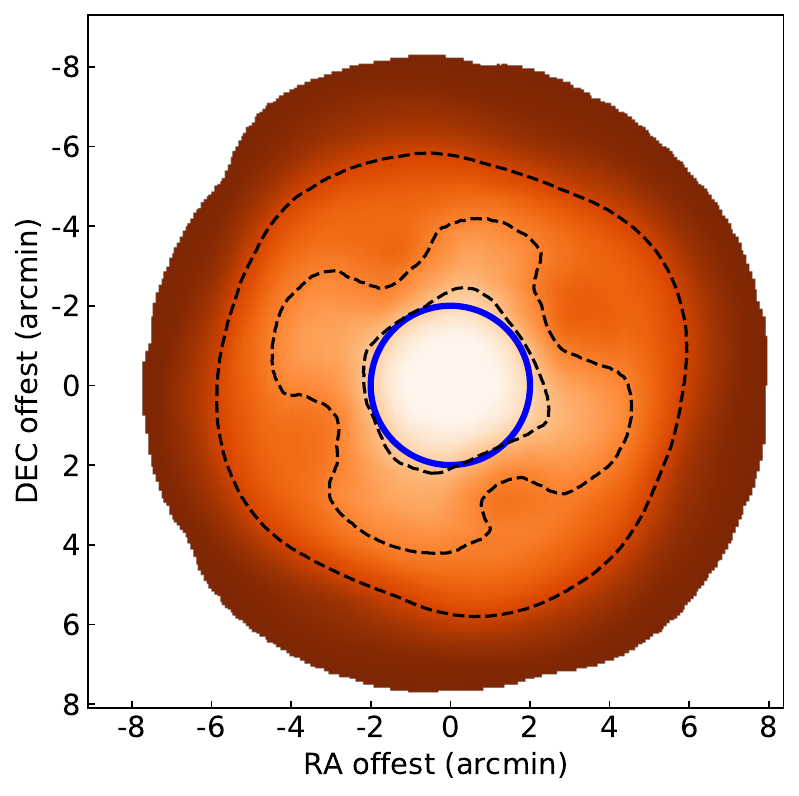}
\caption{Example SCUBA2 exposure time map. The contours denote the 20, 40, 60\% exposure time levels compared to the maximum level on the map. The blue circle region has a diameter of 4\arcmin\, and marks the region with uniform and low RMS for SMG overdensity searches.}
\label{figure_time}
\end{figure}

\begin{figure*}
\centering
\includegraphics[width = \linewidth]{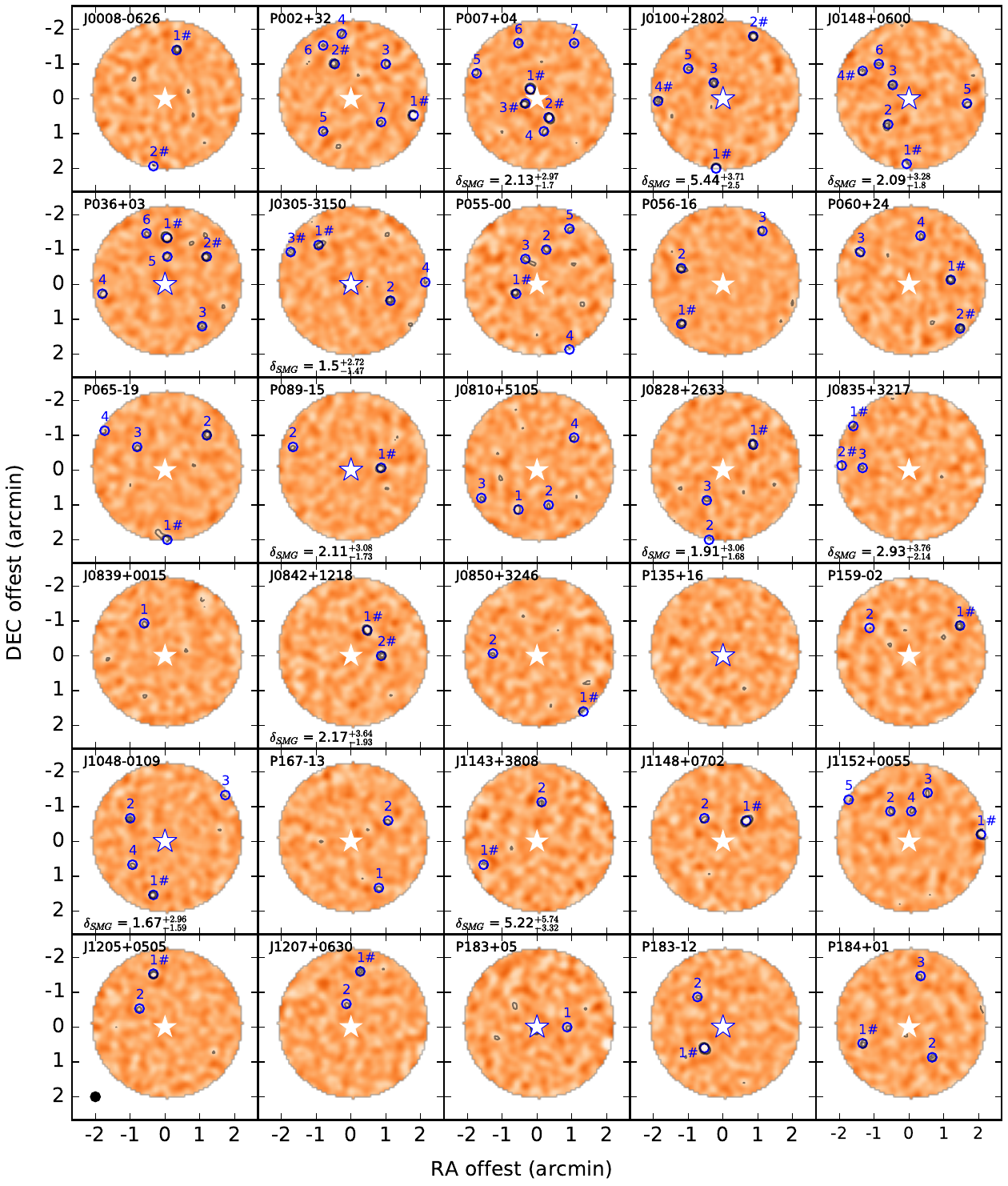}
\caption{JCMT-SCUBA2 850\um\, images of the surrounding regions around 54 $z \sim 6$ quasar with a beam size of FWHM = 15\arcsec.
The selected region is within 2.23 arcmins from the quasar optical position (which covers typical protocluster scales at $z \sim 6$) and has one sigma uncertainty of $<$ 1.5 mJy beam$^{-1}$ for each pixel. The contour levels are +3, +4, +5, +6, +7$\times$1.5 mJy beam$^{-1}$ for each map. The star indicates the quasar optical/NIR position. The star with the blue edge is the sub-mm detected quasars. The blue circles indicate the selected SMGs (with S/N$>$3.5 at 850um or 450um, see Table~\ref{table2}).
The sources marked with `\#' are the SMGs used in the overdensity analysis with deboosted fluxes $>$ 5.5 mJy.
}
\label{figure_1}
\end{figure*}

\begin{figure*}
\centering
\includegraphics[width = \linewidth]{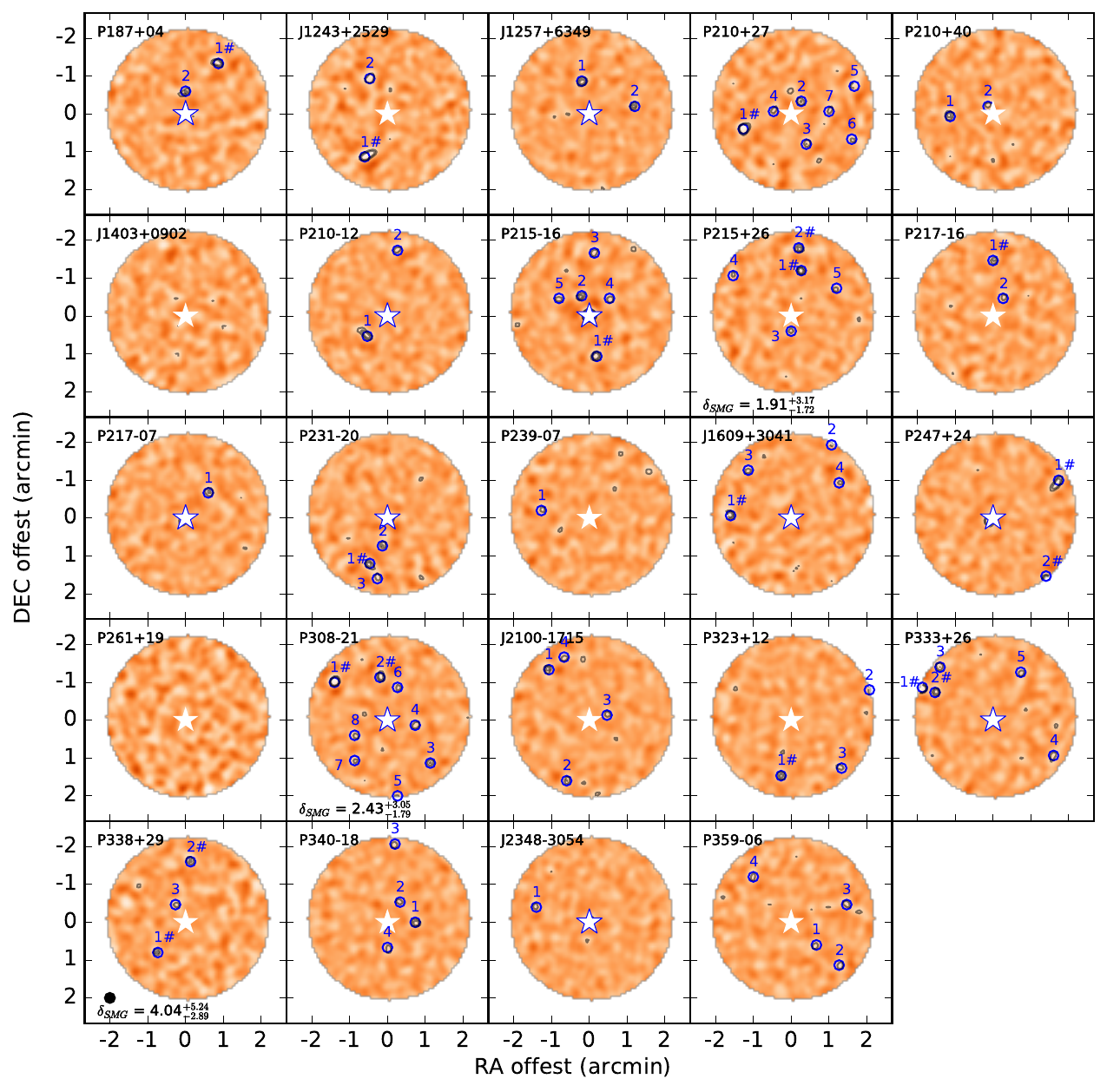}
\caption{-Continued}
\end{figure*}


\section{Source extraction and Number counts}\label{section 3}
\subsection{Source extraction}\label{section 3.1}
The SHERRY survey fields are centered on the targeted $z\sim6$ quasars.
The expected protocluster size is about $\sim$ 20 cMpc along the side, and the core of the protocluster is 2 - 3 cMpc \citep{Overzier2009,Chiang2017}.
For each map, we use the (square root of the) exposure time as a proxy for noise, which is as a function of radius (Figure\,\ref{figure_time}). Note the actual noise map is produced from the pipeline, including other noise, e.g., the instrument and atmospheric noise.
Considering that the noise increases rapidly at a large radius of the Daisy maps, we searched for overdensity within the central $\sim$2 arcmins around the quasar. This region corresponds to a scale of $\sim$10 cMpc in diameter, which can trace a typical protocluster \citep{Chiang2013}.
Furthermore, we mask the region where the RMS noise level is $>$ 1.5 mJy beam$^{-1}$ based on the error map to ensure the best/uniform sensitivity in the SCUBA-2 images. The error difference between the center and the edge is less than $\sim15\%$. The source extraction is processed in this effective region, shown in Figure \ref{figure_1}.

To extract the sources in the maps, we adopted the `top-down' algorithm to find the source with the highest SNR in the two bands maps for each quasar field (e.g., \citealt{ChenCC2013,Geach2017}).
SMGs at high redshifts are unresolved in SCUBA-2 observations; thus, the flux density of the source is the peak value of the pixels in the beam.
We first identified the pixel's position with the highest SNR for each source.
Then we subtracted a scaled point spread function (PSF) from the image at the source position.
We generated the Gaussian PSF model with an FWHM of 14.9 arcsec at 850 \um\, map and 7.9 arcsec at 450 \um\,.
During each iteration, we recorded the coordinates, flux densities, noise values, and SNRs of each $>3.0\sigma$ detection of candidate sources. 
To create a unique source list, we removed multiple detections if the secondary peaks were within one beam area of a higher SNR one.
We iterated this process until there were no more sources with SNR of greater significance than the threshold in the image.

After identifying the $>$ 3.0$\sigma$ candidate sources in the two band maps, we cross-checked the sources between 850\um\, and 450\um\, maps to find the counterpart in the radius of 15\arcsec\, (the beam size at 850\um\, map).
The final catalog of submm source candidates is generated as those satisfying $>$ 3.5$\sigma$ at least  in one band. The flux density of its counterpart on another band is also listed in the catalog, if it has.
The 850\um\, images of each of the 54 quasar fields are shown in Figure \ref{figure_1}. The final SMGs selected as described above are marked as blue circles, and listed in Table~\ref{table2} of Appendix \ref{SMG}.
We also carried out MCMC deboosted process for each individual source’s flux for repeating 50000 times, and calculated the deboosted flux and its error.


\begin{figure*}
\includegraphics[width = \linewidth]{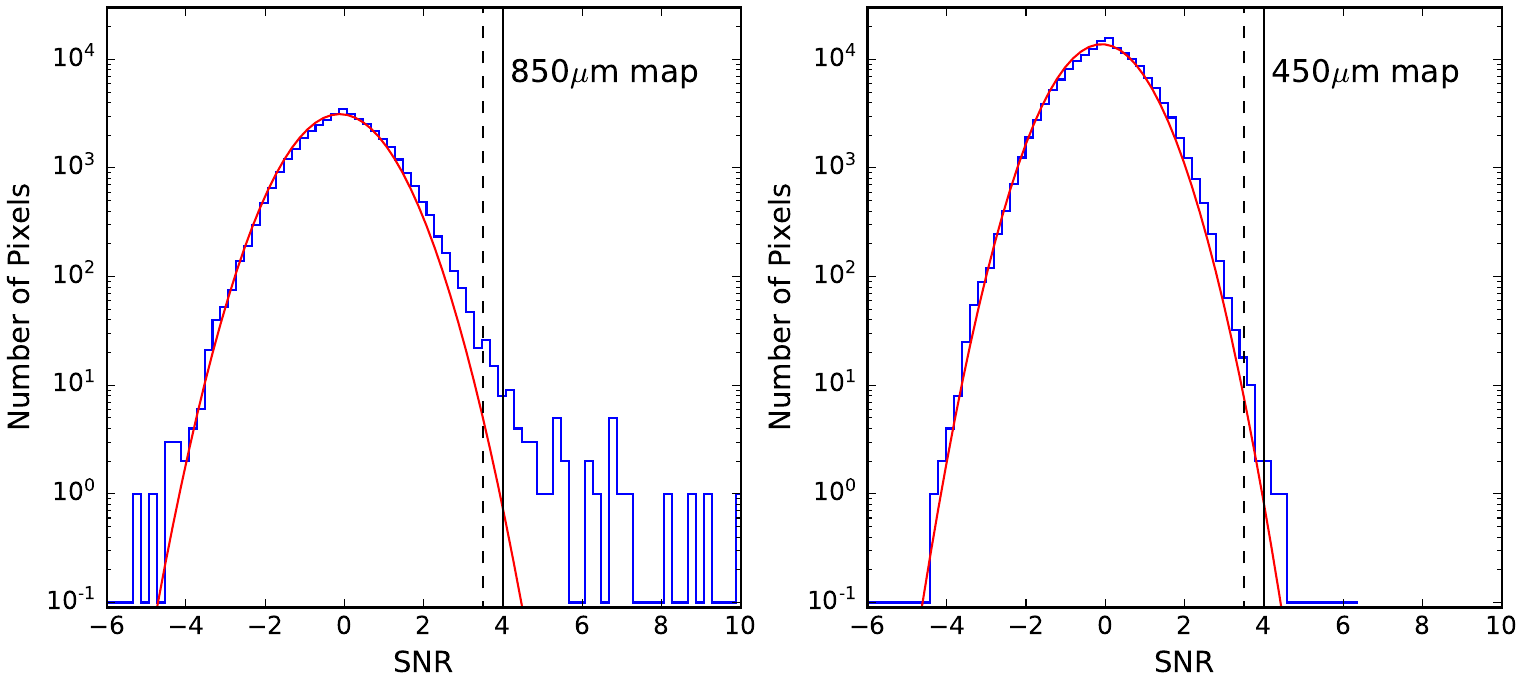}
\centering
\caption{The S/N histograms of the signal maps in the field of quasar SDSS J0100+2802 ($z=6.3258\pm0.0010$) in the 850\um\, and 450\um\, bands. The dashed and solid lines indicate the 3.5$\sigma$ and 4$\sigma$. The red line represents the Gaussian fitting of the noise statistics.}
\label{figure_5_pixels}
\end{figure*}
\begin{figure*}
\centering
\includegraphics[width = \linewidth]{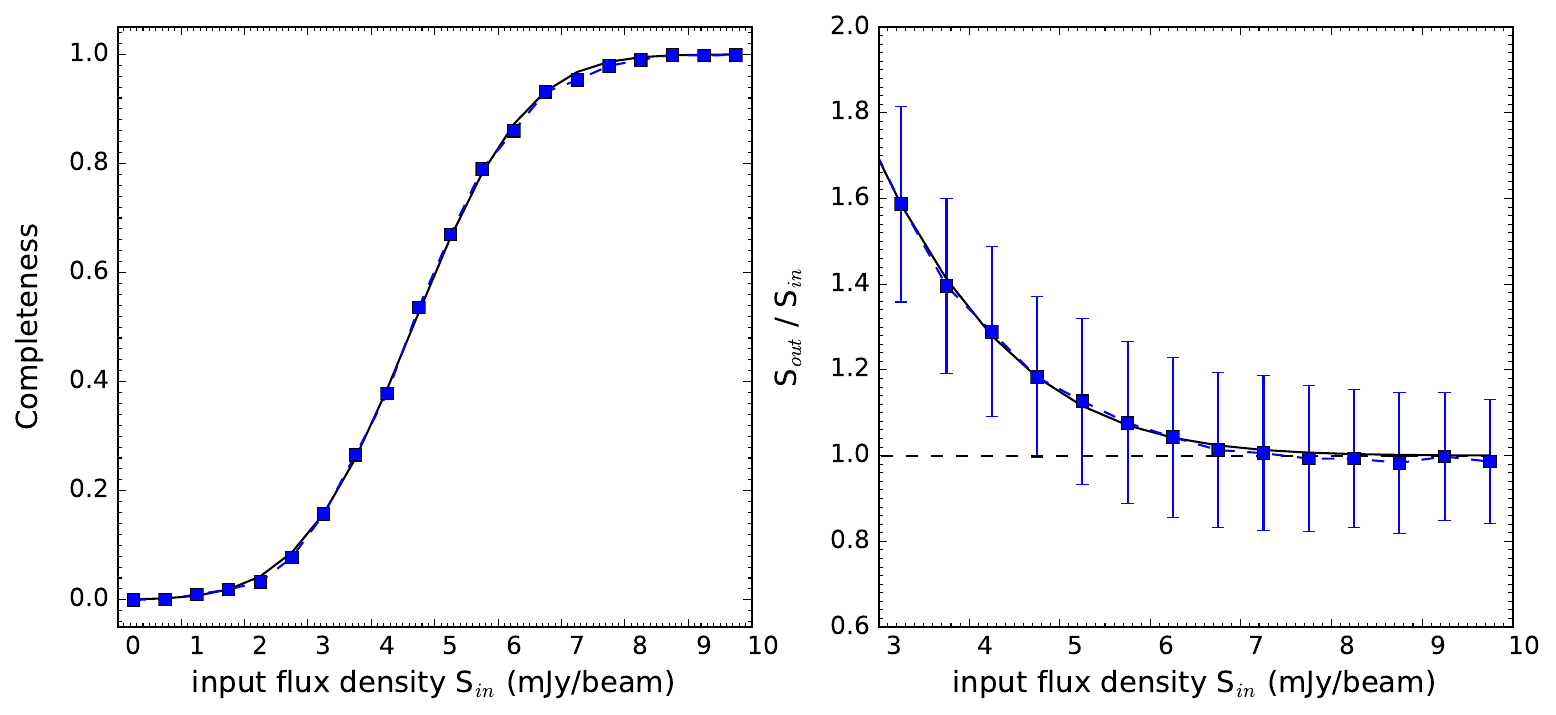}
\caption{Example completeness and flux boosting estimates based on Monte Carlo source insertion. Left: Completeness as a function of input flux (S$_{in}$). The solid curve represents the best-fit function of \emph{f(S$_{in}$) = [1+erf((S$_{in}$-A)/B)]/2} with best fitting parameters give in the text. Right: The ratio between input flux density and output flux density as a function of input flux. Error bars show 1$\sigma$ of 5000 trials. Solid curve represents the best-fit function of \emph{f(S$_{in}$) =1+Aexp(-BS$_{in} ^C$)}. \citep{Hatsukade2016}.}
\label{fig:completeness and boosting}
\end{figure*}

\subsection{Number counts}
In this section, we determine the pure galaxy number counts in the quasar fields. To do so, we first need to estimate the number of spurious sources contaminating the counts and estimate completeness levels, particularly at low SNR.
These completeness and flux deboosting calculations are all processed for each field individually. In this section, we use SDSS J0100+2802 as an example to show how it works.
Note the number counts result in this paper is only based on 850 \um\, maps due to its better sensitivity.

\subsubsection{Spurious source}
We examined the negative noise statistics to calculate the contamination rate of false detections.
For example, we plot the S/N histograms of the pixels in the effective area of the signal maps at quasar nearby field for SDSS J0100+2802 ($z=6.3258\pm0.0010$) in 850\um\, and 450\um\, bands, in Figure \ref{figure_5_pixels}. The red line represents the Gaussian fitting of the noise statistics.
The excess above this line on the positive side represents true detections assuming that the noise is symmetric.
On the other hand, the negative tail is from the spurious sources due to the matched-filter PSF \citep{Chapin2013}. We need to estimate the distribution of negative to obtain the pure number counts.

We performed this check on each field individually. We first constructed an inverted flux map and then applied the source extraction procedure described in Section\,\ref{section 3.1} to them. Nine detections at $\sigma>4$ were found in the inverted maps at 850\um\, band. We estimated the false detection rate above $4\sigma$ significance to be 11.4\% (9/79). We also estimated the false detection rate at $3<\sigma<4$ to be 57.6\% (57/99).
Considering the effects of spurious sources, we calculate the number counts with the flux bin beginning from 4$\sigma$. Then the impact of the spurious source will significantly diminish, and the conclusions below are not affected.


\subsubsection{Completeness} \label{sec:Completeness}
To describe how we make some corrections to calculate the number counts, we still use the field of quasar SDSS J0100+2802 ($z=6.3258\pm0.0010$) as an example in the following.
The completeness is the rate at which a source is expected to be detected in the image \citep{Hatsukade2016}. We used the Monte Carlo simulation approach to estimate completeness. The completeness test is only in the central best/uniform-sensitivity regions.
We injected artificial sources of $0-10$\,mJy/beam at random positions in the maps $>$ 8.0\arcsec\, away from $\geq$ 3$\sigma$ peaks to avoid the contribution from nearby sources. The flux bin step is 0.5 mJy for 850 \um\, map.
Then we ran the source extraction on the maps to see if we could recover the injected sources.
When the output source is above 3$\sigma$, the source is considered to be recovered. This procedure is repeated 5000 times for each flux bin.
We fit the completeness as a function of
\emph{f(S$_{in}$)={\rm [1 +  erf((}S$_{in}$ - A)/B{\rm )]/ 2}}
\citep{Hatsukade2016}, as shown in Figure \ref{fig:completeness and boosting}. Here S$_{in}$ is the input flux density.
The best-fitted parameters are $A = 4.67\pm0.01$, $B = 2.05\pm0.02$.
Based on this Monte-Carlo experiment, we estimate a 50\% completeness level for 4.7 mJy point sources and 80\% for 5.9 mJy point sources in 850\um\, map.

\subsubsection{flux boosting}
With the Monte Carlo simulations, we also considered the effect that flux densities of low S/N sources are boosted \citep{Murdoch1973,Hatsukade2016}. More specifically, we calculated the ratio between the flux of the recovered sources and those of the injected artificial sources. We then estimate the mean ratio in 0.5 mJy flux bins.
To illustrate this process, we plot the mean flux boosting value for the field of J0100 in Figure~\ref{fig:completeness and boosting} (right).
The flux boosting is more pronounced for low S/N sources. We fit the ratio between input flux density and output flux density as a function of \emph{f(S$_{in}$) =1+Aexp(-BS$_{in} ^C$)}\citep{Hatsukade2016}.
The best fitted parameters are $A = 1.74\pm0.27$, $B = 0.11\pm0.04$ and $C = 1.95\pm0.22$.
At S/N = 4 (input flux = 4.8 mJy), the estimated mean flux boosting factor is 1.19 at 850 \um\, map. This is consistent with the results of Monte Carlo simulations in previous SCUBA studies \citep{Eales2000,Scott2002,Cowie2002,Casey2013,ChenCC2013}, such as the boosting factor of 1.25 at 850\um\, map in \citet{Cowie2002}.
Potential contamination for the SMG number counts is the gravitational lensing effects of foreground objects (e.g., \citealt{Chen2011}). If some of the detected submillimeter sources are lensed background sources, they may increase the uncertainties of the source positions and also boost the measured fluxes. \citep{Knudsen2008}

\begin{figure*}
\centering
\includegraphics[width = \linewidth]{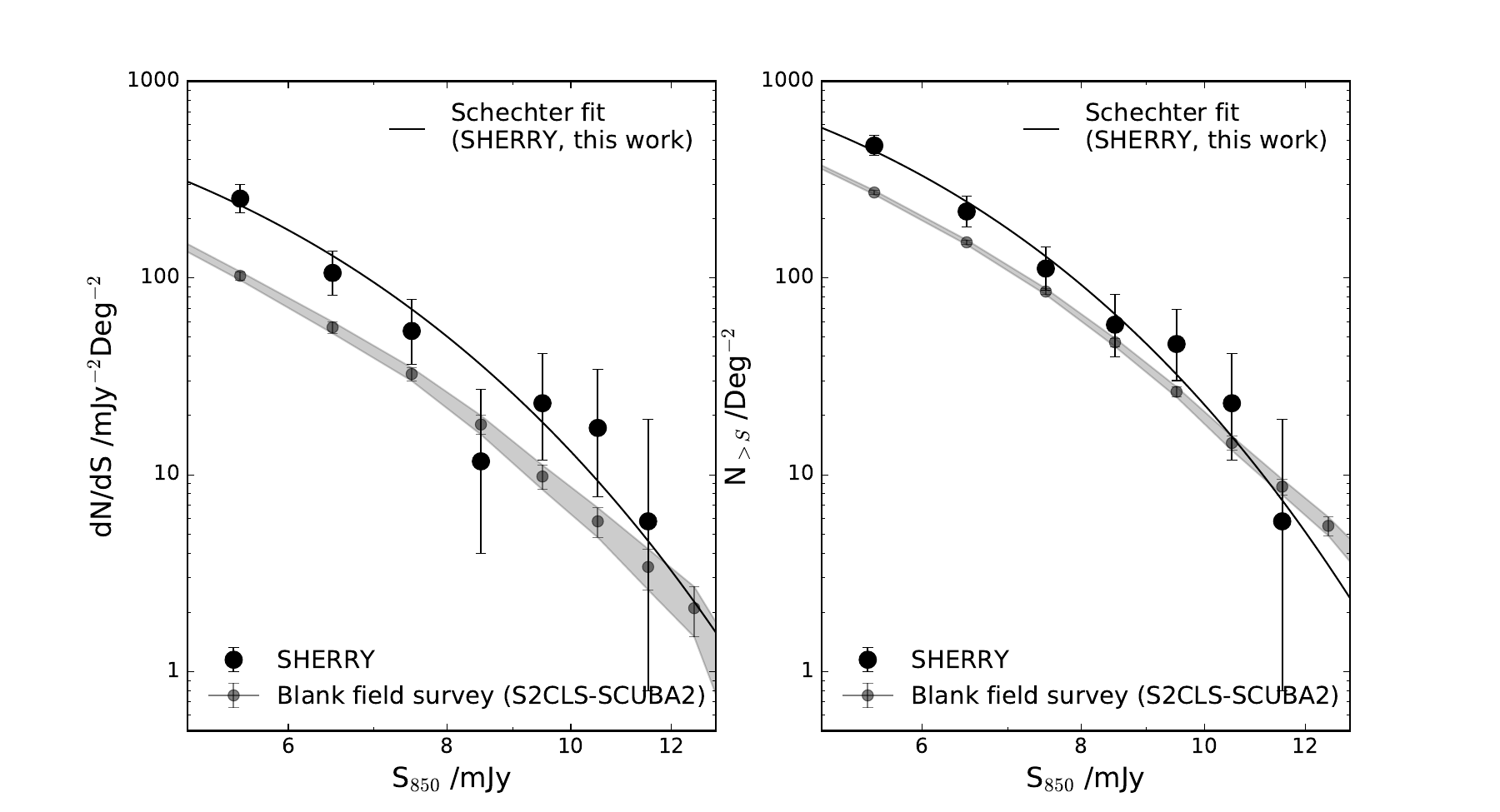}
\caption{
The combined differential and cumulative number counts at 850\um\, from all of our source fields. The filled black points and black lines indicate our data and Schechter fitting, respectively.
The gray line shows the blank-field number counts of SMGs in the S2CLS field in \cite{Geach2017}.
For clarity, we only show the error bars of S2CLS and our data from Poissonian statistics \citep{Gehrels1986}.
Points in this work have been corrected for flux boosting and incompleteness (see figure~\ref{fig:completeness and boosting}).}
\label{figure6}
\end{figure*}

\begin{deluxetable}{ccccc}
\tablecaption{The combined differential and cumulative number counts at 850\um\label{table1}}
\tablehead{
\colhead{S$_{860}$}   &\colhead{dN} &\colhead{dN$_{corr}$}  & \colhead{dN/dS}           & \colhead{N($>$S)} \\
\colhead{(mJy)} &             &                 & \colhead{(mJy$^{-1}$deg$^{-2}$)}                         & \colhead{(deg$^{-2}$)}
}
\startdata
5.5 & 29 & 42.2 $_{ -6.5 }^{+ 7.6 }(\pm7.4)$ & 243.5 $_{ -37.4 }^{+ 43.6 }(\pm40.1)$ & 461.4 $_{ -51.5 }^{+ 57.6 }(\pm53.4)$ \\
6.5 & 16 & 18.4 $_{ -4.3 }^{+ 5.4 }(\pm4.5)$ & 106.1 $_{ -24.6 }^{+ 31.0 }(\pm24.9)$ & 217.9 $_{ -35.4 }^{+ 41.6 }(\pm33.9)$ \\
7.5 & 9 & 9.3   $_{ -3.0 }^{+ 4.2 }(\pm2.6)$ & 53.9  $_{ -17.4 }^{+ 24.1 }(\pm15.4)$ & 111.8 $_{ -25.2 }^{+ 31.7 }(\pm22.1)$ \\
8.5 & 2 & 2.0   $_{ -1.3 }^{+ 2.7 }(\pm1.4)$ & 11.7  $_{ -7.7 }^{+ 15.4 } (\pm8.3 )$ & 57.9  $_{ -18.1 }^{+ 24.7 }(\pm16.4)$ \\
9.5 & 4 & 4.0   $_{ -1.9 }^{+ 3.2 }(\pm1.9)$ & 23.1  $_{ -11.2 }^{+ 18.4 }(\pm11.2)$ & 46.2  $_{ -16.1 }^{+ 22.9 }(\pm14.3)$ \\
10.5 & 3 & 3.0  $_{ -1.7 }^{+ 2.9 }(\pm1.7)$ & 17.3  $_{ -9.6 }^{+ 17.0 } (\pm9.6 )$ & 23.1  $_{ -11.2 }^{+ 18.4 }(\pm11.5)$ \\
11.5 & 1 & 1.0  $_{ -0.9 }^{+ 2.3 }(\pm1.0)$ & 5.8   $_{ -5.0 }^{+ 13.4 } (\pm5.8 )$ & 5.8   $_{ -5.0 }^{+ 13.4 } (\pm5.8 )$ \\
\enddata
\tablecomments{\\
S is the bin central flux density, and the bin width is one mJy. The first set of uncertainties is Poissonian errors. And the second reflects a bootstrap analysis error that is the standard deviation of counts in each bin after 1000 realizations of the random sampling. These uncertainties are of comparable magnitude to the Poisson errors.
The combined area includes all 54 quasar nearby fields as a statistical field with the size of $\sim$ 620 arcmin$^2$, corresponding to $\sim$ 3700 cMpc$^2$ ($\sim$ 76 pMpc$^2$) at $z \sim 6$.
}
\end{deluxetable}

\section{Results and Discussion}\label{section 4}
\subsection{Differential and cumulative number counts}
To obtain an unbiased estimate of the number counts for each SMG candidate, we statistically correct the observed flux density using the Monte Carlo simulation for each field, see Figure \ref{fig:completeness and boosting}. Then, the deboosted flux density is amended for completeness, as described in Section~\ref{sec:Completeness}. Note we also removed the quasar itself in the analysis of number counts.

There are two number counts to describe the source abundance: the cumulative number counts and the differential number counts.
To study the general environment of $z \sim 6$ quasars, we built a combined area to calculate the differential number counts, which have independent errors for each count measurement, making it straightforward to do fitting. The combined area includes all 54 quasar nearby fields and totals $\sim$ 620 arcmin$^2$, corresponding to $\sim$ 3700 cMpc$^2$ ($\sim$ 76 pMpc$^2$) at $z \sim 6$. It represents the overall environmental property of quasar fields at $z \sim 6$.
Another number count is the cumulative number count. It applies to the small sample size with a visual underlying counts shape. This method can be used in the analysis of individual quasar nearby fields. 

The differential number counts of the combined area are calculated by dividing each $>$ 3.5$\sigma$ detection by the total combined area and adding up all measurements lying with each flux bin, then dividing by the flux width. Following previous studies, we fit the differential number counts with a Schechter function of the form \citep{Casey2013,Geach2017}:
\begin{equation}\label{dNdS}
\begin{split}
& \frac {dN}{dS} = \left(\frac {N_0}{S_0}\right) \left(\frac {S}{S_0}\right)^\alpha {\rm exp} \left(-\frac {S}{S_0}\right)
\end{split}
\end{equation}
We find the best fit parameters are $N_0 \sim 3000 \rm deg^{-2}$, $S_0 \sim 1.2$ mJy and $\alpha \sim$ 1.5.
We show the corrected differential and cumulative number counts for the effective area of the 54 quasars, together with the fit results in Figure \ref{figure6}. We list the corrected data points in Table \ref{table1}.
Note that our fitting is restricted to sources brighter than 5.5 mJy.
The first set of uncertainties is Poisson errors \citep{Gehrels1986}. And the second reflects the field variance that we used a bootstrapping approach.
The individual field was selected randomly as many times as the size of a given sample, allowing for replacements, to create a new sample to calculate the counts. And each source is also corrected deboosting and completeness. This process was repeated 1000 times, and then we fitted a normal distribution to these 1000 realizations. The standard deviation of this distribution was the field variance.
These variances are of comparable magnitude to the Poisson errors.

\begin{figure*}
\centering
\includegraphics[width = 12cm]{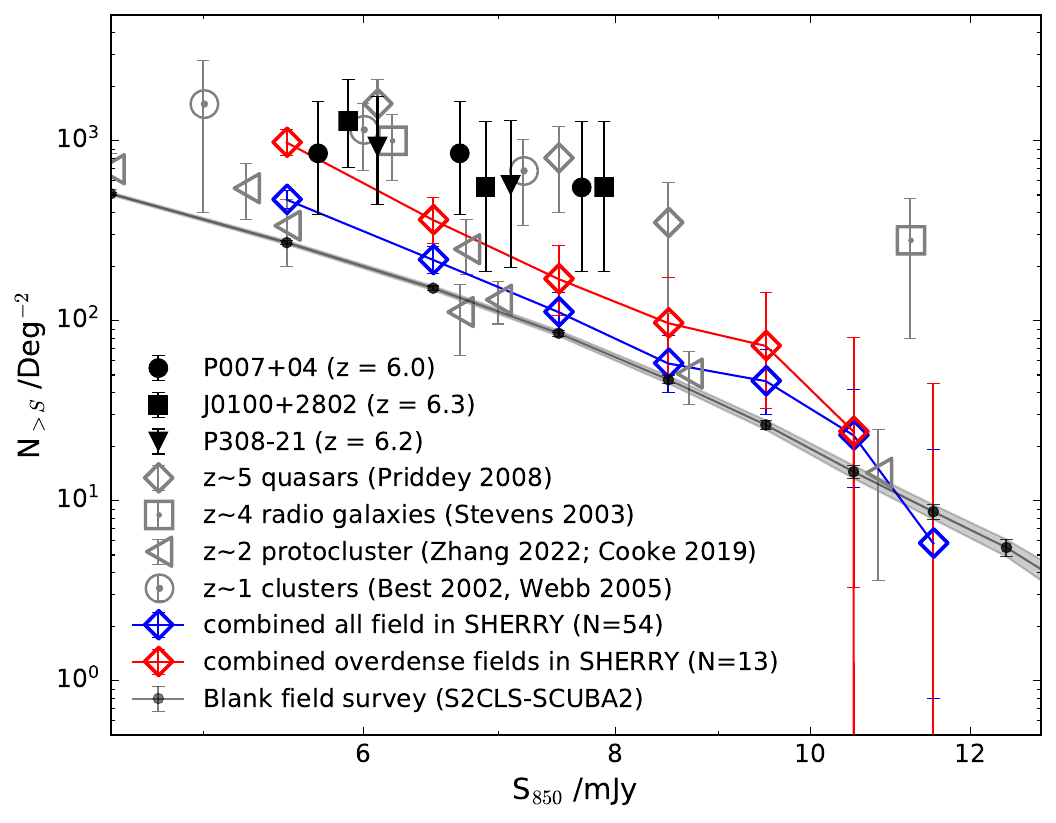}
\caption{Cumulative number counts of SMGs, in fields centered on (i) luminous high-redshift AGN: z $\sim$ 5 optically selected quasars (grey diamonds; \citealt{Priddey2008}); z $\sim$ 4 radio galaxies (grey squares; \citealt{Stevens2003}); (ii) z $\sim$ 2 protoclusters (grey triangles; \citealt{ZhangYuheng2022,Cooke2019}); (iii) z $\sim$ 1 clusters (grey circles; \citealt{Best2002,Webb2005}) and (iv) $z \sim 6$ optically selected quasars (this paper, black points and red/blue diamonds).
The black line shows the blank-field number counts of SMGs in the previous survey, i.e., SCUBA2 survey for S2CLS field in \cite{Geach2017}.
The blue line is calculated from our survey's combined areas of all 54 quasars ($\sim$ 620 arcmin$^2$); the red line is from the combined areas of 13 overdense fields around the quasars ($\sim$ 150 arcmin$^2$).
The black-filled points indicate the three most overdense fields selected from our sample at $z \sim 6$.
Points in this work have been corrected for flux boosting and incompleteness.}
\label{figure_7}
\end{figure*}

We then compared our results with the previous blank field survey - SCUBA2 Cosmology Legacy Survey (S2CLS, $\sim$5 deg$^2$) at 850 \um.
To a $4\sigma$ detection ($\sim$ 5.5 mJy), the SMGs overdensity ({$\delta_{\rm{SMG}}=\rho_{\rm{SMG}}/<\rho_{\rm{SMG}}>-1$}) is $0.68^{+0.21}_{-0.19}$($\pm0.19$) in the combined field of 0.17 degree$^2$, exceeding ($\sim$ 1.68 $\times$) the source counts of blank sky area. And to a $5\sigma$ detection, the SMG overdensity is a factor of $0.40^{+0.27}_{-0.23}$($\pm0.24$).
If the galaxy excess is mainly due to overdensities at a narrow redshift range around $z \sim 6$, the SMG overdensity near quasar redshifts must be higher.


To investigate which quasars reside in an overdense environment and search for protoclusters at $z \sim 6$, we also calculated the completeness and boosting corrected SMG cumulative number counts N($>$S) individually for each quasar field. We compared the results with the number counts in the blank field and gave a rough definition of `overdensity'. The lower limit of the SMGs cumulative number counts to 5.5 mJy ($\sim$ 4$\sigma$) is above the value in the blank field. With this criterion, 13 quasar fields show clear sub-mm overdensities with the overdensity of $\delta_{SMG} \sim $ 1.5 - 5.4. We label their overdensity in Figure \ref{figure_1}.
Here the calculation of the overdensity is based on the sources with deboosted fluxes brighter than 5.5 mJy. There are 55 sources from the whole survey (marked with `\#' in Figure~\ref{figure_1}). 24 of these lie in the 13 over-densities fields, with four of the over-densities comprise just one source above this flux limit.

We also combined the overdense fields around these 13 quasars and compared them with the blank field and those around other lower redshift galaxies.
For comparison we show the cumulative number counts of SMGs, in fields centered on (i) luminous high-redshift AGN: z $\sim$ 5 optically selected quasars \citep{Priddey2008}; z $\sim$ 4 radio galaxies \citep{Stevens2003}; (ii) z $\sim$ 2 protoclusters \citep{ZhangYuheng2022}; (iii) z $\sim$ 1 clusters \citep{Best2002,Webb2005} and (iv) $z \sim 6$ optically selected quasars (in this paper) in Figure~\ref{figure_7}. The black line shows the blank field number counts of SMGs from the SCUBA2 survey of the S2CLS field in \citet{Geach2017}.
The combined field of these 13 quasars exceeds the blank field counts with the overdensity to 5.5 mJy of \dsmg $\sim$ $2.46^{+0.64}_{-0.55}$($\pm0.25$) in the regions of $\sim$ 150 arcmin$^2$.
It is consistent with the lower redshift cluster and the surrounding regions of quasars and radio galaxies. But to the bright end, the excess is not significant, e.g., to 7.5 mJy, the overdensity decreases to 1.00$^{+1.08}_{-0.74}$($\pm0.73$).
Compared with the lower redshift galaxies/quasar fields or cluster fields, the combined overdense area shows a lower overdensity. That is because the data from the literature are case studies that intentionally target overdense regions (including significant selected bias). Still, our data points represent the average environmental properties around $z \sim 6$ quasars, and overdensity is diluted.

Therefore, considering the abundant SMGs candidates in the faint-end, we select the fields where the lower limit of the SMGs cumulative number counts to 6.5 mJy ($\sim$ 5$\sigma$) is above the value in the blank field. There are three (3/13) $z \sim 6$ fields that show a significant excess over the blank field: P007+04 ($z=6.0$), J0100+2801 ($z=6.3$) and P308-21 ($z=6.2$), shown in Figure \ref{figure_7}.
They are more likely similar to the lower redshift overdense fields/clusters with the overdensity of 2.1 - 5.4 at 5.5 mJy.
However, \citealt{Champagne2018} analyzed 1.2 mm ALMA dust continuum maps for 35 bright quasars at $z\sim6$. They reported there is no evidence of 0.1 - 1mJy millimeter continuum sources in the environments of P007+04 ($z = 6.0$) and P308-21 ($z = 6.2$). The overdensity is -0.07$\pm$0.56, which implies no detected overdensity of dusty galaxies around these quasars.
This conclusion is based on a small ALMA field of view of about 25\arcsec\,, corresponding to an angular diameter distance of $\sim$ 1 cMpc at $z \sim 6$. It could miss the overdensity on scales larger than one cMpc.

\subsection{Correlation with LAE and LBG at $z \sim 6$}
Some previous works used other tracers, like LAEs and LBGs, to probe the large-scale structures.
But the question of whether the galaxy environment traced by LAEs or LBGs of $z \sim 6$ quasar is overdense or not is still far from clear.
\citet{Morselli2014} showed that the environment traced by LBGs around four $z \sim 6$ quasars is overdense in a wide area of $\sim$ 575 arcmin$^2$, $\sim$3100 cMpc$^2$. Conversely, \citet{Willott2005} reported three SDSS quasars at $z \sim 6$ that have no clear signs for LBG overdensity in an area of $\sim$ 30 arcmin$^2$ ($\sim$170 cMpc$^2$ at $z \sim 6$).

At $z = 2 - 6$, some observations reported that SMG brightness appears to be correlated with the high-density region of LAEs and LBGs (e.g. \citealt{Pope2006,Tamura2009,Matsuda2011,Hodge2013,Umehata2017,Harikane2019}). 
Here we compare some individual sources that have previous environment studies traced by LAEs and LBGs on a similar scale.
\citet{Becker2018} reported that J0148+0600 field has a highly significant deficit of LAEs. It is roughly one-quarter of background density with ten cMpc (similar size in our survey). But in our millimeter survey, the environment of J0148+0600 has an overdensity of $\sim$ 2.1 ($\sim$ 3 $\times$ the background field).

Conversely, \citet{Mazzucchelli2017b} reported no evidence for an overdensity of LAEs in an area of $\sim$ 37 arcmin$^2$ centered on the quasar PSO J215-16.
We also find no evidence of SMG overdensity in the field of PSO J215-16 in our SHERRY survey.
At a similar scale, the different findings may be ascribed to different tracers.
These LAEs or LBGs trace the young, dust-poor star-forming galaxies. Nevertheless, some observations and cosmological simulations indicate that galaxy assembly at high redshift around the quasar might be characterized by galaxies with elevated dust and molecular gas content. \citep[e.g.][]{DeRosa2011,Yajima2015}.
The SMGs trace massive, metal-enriched, and highly obscured objects, which may be a more suited tracer for high redshift overdensities. SMGs, therefore, give us a new perspective on the studies of galaxy environment at high redshift.

\subsection{Massive SMGs in proto-clusters}
Our SHERRY survey covers 54 $z \sim 6$ quasars (20/54 detected at the sub-mm band, see \citealt{LiQ2020}). We identified 170 SMGs ($>3.5\sigma$) at 450 and 850 \um, in an area of $\sim$620 arcmin$^2$. SHERRY represents the first systematic study to probe the environment of $z\sim6$ quasars in the sub-mm bands.
For these SMGs lie $z \sim 6$, their FIR luminosities are 2.2 - 6.4 $\times$ 10$^{12} L_{\odot}$, and their star formation rates are $\sim$ 400 - 1200 $M_{\odot}\, yr^{-1}$ (assuming a modified black body SEDs with a typical SMG dust temperature of 35 $K$, an emissivity index of $\beta$ = 1.6 and a Salpeter IMF).

We found that 13/54 fields around $z \sim 6$ quasars exhibit excess above SMG counts compared to a blank field. However, we caution that the other 41/54 quasars could also reside in the overdense fields, because our survey is not sensitive to the fainter SMGs with $L<$ 2.0 $\times$ 10$^{12} L_{\odot}$.
Furthermore, even in an actual dense region of galaxies, the expected number of massive galaxies will not be very large based on the luminosity function.

On the other hand, our SHERRY survey does not provide an accurate redshift determination. Therefore, the SMG overdensity may be diluted over the large cosmological volume.
\citet{Meyer2022} presented ALMA observations for three quasar-SMG fields (P231-20 at $z = 6.59$, P308-21 $z = 6.24$ and J0305-3150 $z = 6.61$) selected from our SCUBA2 survey to search for [C{\sc ii}] emission.
12 out of 17 SCUBA2-selected sources are detected with ALMA in the continuum. ALMA detections have offset from the center of pointing due to the large beam of SCUBA2 imaging (13\arcsec\, at 850\um\,, e.g., of the order of ALMA field of view, Figure~1 in \citealt{Meyer2022}).
They also search for the \cii\, emission lines in six of the continuum-detected sources: four ones in the field of P231-20 ($>5\sigma$) and two ones around P308-21 (3-4$\sigma$).
In the field of PJ231-20, they confirmed a foreground cluster at $z\sim2.4$.
But unfortunately, none of the sources have a photo-$z$ consistent with the background $z\sim6$ quasars. In the future, more high-resolution imaging and spectroscopic observations are required for these JCMT-selected protocluster candidates.
If we confirm that a significant fraction of the SCUBA-detected galaxies are indeed at the redshift of the quasars (e.g., P007+04, J0100+2802, etc.), we will have striking evidence that some high-$z$ quasars reside in highly overdense regions.

\section{Summary}\label{section 5}
In this paper, we have presented a SCUBA-2 sub-mm survey `SHERRY' of 54 $z \sim 6$ QSOs to study their environment and to search for galaxy protoclusters at the early epoch.
The large field of view of JCMT enabled us to identify 170 SMGs ($>3.5\sigma$) at 450 and 850 \um\,. SMGs are selected within a distance of $\sim$2 arcmins away from the central quasars and have the RMS noise level of $\sim$ 1.2 mJy beam$^{-1}$ in the 850 \um\, band. This region corresponds to 10 cMpc in diameter, which can trace the typical galaxy protocluster \citep{Chiang2013}.
If the SMGs are near the quasar redshifts, their FIR luminosities are 2.2 - 6.4 $\times$ 10$^{12} L_{\odot}$ and star formation rates are $\sim$ 400 - 1200 $M_{\odot}\, yr^{-1}$.
The SMGs discovered in SHERRY provide us with essential candidates for follow-up spectroscopic observations (e.g., [C$\rm _{II}$]) to secure redshifts and study the first massive large-scale structures in the universe.

We also calculated the SMGs differential and cumulative number counts after correcting for completeness and flux boosting in a combined area of $\sim$ 620 arcmin$^2$, which corresponds to $\sim$ 3700 cMpc$^2$ ($\sim$ 76 pMpc$^2$) at $z \sim 6$.
To a $4\sigma$ detection (at $\sim$ 5.5 mJy at 850\um\,), the SMGs overdensity is $0.68^{+0.21}_{-0.19}$($\pm0.19$), exceeding ($\sim$ 1.68 $\times$) the source counts of blank sky surveys (S2CLS; \citealt{Geach2017}).
We found 13/54 quasars show SMG overdensities (at $\sim$ 5.5 mJy) in their fields, and the others do not.
The combined field of these 13 quasars exceeds the blank field counts with the overdensity to 5.5 mJy of \dsmg $\sim$ $2.46^{+0.64}_{-0.55}$($\pm0.25$) in the regions of $\sim$ 150 arcmin$^2$; but to the bright end (e.g., 7.5 mJy), the overdensity is not significant.
Among them, three (3/13) $z \sim 6$ fields show a significant excess over the blank field, with the overdensity of 2.1 - 5.4 at 5.5 mJy.

We also compared our work and some previous environment studies traced by LAEs and LBGs on a similar scale.
These LAEs or LBGs trace the young, dust-poor star-forming galaxies, while the SMGs trace massive, metal-enriched, and highly obscured objects.
They may be correlated in the high-density region; therefore, more follow-up spectroscopic works to search for LAEs with JWST, HST, VLT, and ALMA. are required for these fields at a large scale.


\vspace{1cm}

\begin{acknowledgments}

This work was supported by the National Science Foundation of China (NSFC, 11721303, 11991052), the National Key R\&D Program of China (2016YFA0400703), and the NSFC grants No. 11533001.
R.W. acknowledge the support from the Thousand Youth Talents Program of China, NSFC grants No. 11473004.
X.-B. Wu thanks the support from the National Science Foundation of China (11533001, 11721303).
B.V. acknowledges support from the ERC Advanced Grant 740246 (Cosmic Gas).
The authors wish to recognize and acknowledge the significant cultural role and reverence that the summit of Maunakea has always had within the indigenous Hawaiian community. We are most fortunate to have the opportunity to conduct observations from this mountain.
We are grateful to Paul Hewett, Richard McMahon, Daniel Mortlock, and Stephen Warren, who supplied the spectrum of ULAS J0828+2633.
We also thank our support scientists and telescope schedulers: Harriet Parsons, Mark G. Rawlings, Iain Coulson, Steven Mairs, and Jan Wouterloot, for the JCMT observation and data reduction.
\end{acknowledgments}

\appendix

\section{`SHERRY' - SMG candidates catalog at quasar field at $z \sim 6$}\label{SMG}
Here we present SMGs and SMG candidates in the quasar fields at $z \sim 6$. The full catalog includes 170 SMGs selected from 450 and 850 \um\, images. Column (1) contains the SMG IDs.
SMGs are detected above 3.5$\sigma$ at least in one band, marked as `ObjectShortName-SMG+index'. Columns (4) and (7) are the SMG peak flux density in 850\um\, and 450\um\, bands of SCUBA-2 images; Columns (5) and (8) are signal-to-noise ratio in each band, respectively. Columns (6) and (9) are the deboosted flux density for 850\um\, and 450\um\, detections calculated from the simulation. Column (10) is the angular distance between SMGs and their central quasar.

\begin{longrotatetable}
\begin{deluxetable*}{lllccccccc}
\tablecaption{The catalog of SMGs and its candidates in the quasar nearby fields at $z\sim6$ \label{table2}}
\center
\tabletypesize{\tiny}
\tablehead{
\colhead{Source} &\colhead{RA}&\colhead{DEC       } &\colhead{S$\rm _{\nu,850\mu m}$}   &\colhead{S/N$_{850}$}    &\colhead{S$\rm _{\nu,850\mu m}^{Deboosted}$} &\colhead{S$\rm _{\nu,450\mu m}$} &\colhead{S/N$_{450}$} &\colhead{S$\rm _{\nu,450\mu m}^{Deboosted}$} &\colhead{offset}\\
\colhead{      } &\colhead{  (J2000)  }&\colhead{  (J2000)} &\colhead{ (mJy)} &\colhead{    } &\colhead{   (mJy)    } &\colhead{    (mJy)     } &\colhead{       } &\colhead{(mJy)         } &\colhead{ (arcmin) }     \\
\colhead{  (1) } &\colhead{  (2)    } &\colhead{  (3)   } &\colhead{ (4)} &\colhead{  (5)       } &\colhead{ (6)     } &\colhead{  (7)  } &\colhead{ (8)         } &\colhead{(9)} &\colhead{  (10)    }
}
\startdata
{\bf J0008-0626-SMG1 } & 00h08m24.4s & -06d24m40.5s & 6.92$\pm$1.32 & 5.24 & 6.87$\pm$1.86 & \nodata & \nodata & \nodata & 1.440 \\
{\bf J0008-0626-SMG2 } & 00h08m27.1s & -06d28m00.5s & 7.11$\pm$1.50 & 4.75 & 7.08$\pm$2.02 & 45.15$\pm$12.29 & 3.67 & 43.28$\pm$14.26 & 1.962 \\
\hline
{\bf P002+32-SMG1 } & 00h09m23.3s & +32d51m44.9s & 10.73$\pm$1.45 & 7.39 & 10.82$\pm$2.10 & \nodata & \nodata & \nodata & 2.193 \\
{\bf P002+32-SMG2 } & 00h09m34.1s & +32d53m12.9s & 5.86$\pm$1.25 & 4.70 & 5.65$\pm$1.69 & \nodata & \nodata & \nodata & 1.144 \\
{\bf P002+32-SMG3 } & 00h09m27.1s & +32d53m12.9s & 4.96$\pm$1.34 & 3.70 & 4.48$\pm$1.52 & \nodata & \nodata & \nodata & 1.555 \\
{\bf P002+32-SMG4 } & 00h09m33.1s & +32d54m04.9s & 5.14$\pm$1.43 & 3.60 & 4.72$\pm$1.63 & \nodata & \nodata & \nodata & 1.893 \\
{\bf P002+32-SMG5 } & 00h09m35.6s & +32d51m16.9s & 4.38$\pm$1.22 & 3.59 & 3.67$\pm$1.25 & \nodata & \nodata & \nodata & 1.334 \\
{\bf P002+32-SMG6 } & 00h09m35.6s & +32d53m44.9s & 4.79$\pm$1.34 & 3.58 & 4.23$\pm$1.45 & \nodata & \nodata & \nodata & 1.805 \\
{\bf P002+32-SMG7 } & 00h09m27.7s & +32d51m32.9s & 4.30$\pm$1.21 & 3.54 & 3.56$\pm$1.22 & \nodata & \nodata & \nodata & 1.228 \\
\hline
{\bf P007+04-SMG1 } & 00h28m07.4s & +04d57m41.7s & 10.79$\pm$1.08 & 9.95 & 10.87$\pm$1.78 & \nodata & \nodata & \nodata & 0.334 \\
{\bf P007+04-SMG2 } & 00h28m05.2s & +04d56m53.7s & 7.71$\pm$1.12 & 6.86 & 7.77$\pm$1.79 & \nodata & \nodata & \nodata & 0.630 \\
{\bf P007+04-SMG3 } & 00h28m07.9s & +04d57m17.7s & 6.84$\pm$1.10 & 6.24 & 6.88$\pm$1.76 & \nodata & \nodata & \nodata & 0.360 \\
{\bf P007+04-SMG4 } & 00h28m05.7s & +04d56m29.7s & 4.60$\pm$1.17 & 3.94 & 4.11$\pm$1.33 & \nodata & \nodata & \nodata & 0.955 \\
{\bf P007+04-SMG5 } & 00h28m13.5s & +04d58m09.7s & 5.23$\pm$1.43 & 3.65 & 4.97$\pm$1.71 & \nodata & \nodata & \nodata & 1.888 \\
{\bf P007+04-SMG6 } & 00h28m08.7s & +04d59m01.7s & 4.18$\pm$1.17 & 3.56 & 3.54$\pm$1.21 & \nodata & \nodata & \nodata & 1.687 \\
{\bf P007+04-SMG7 } & 00h28m02.3s & +04d59m01.7s & 4.30$\pm$1.22 & 3.51 & 3.70$\pm$1.28 & \nodata & \nodata & \nodata & 1.925 \\
\hline
{\bf J0100+2802-AGN} & 01h00m12.6s & +28d02m25.8s & 4.09$\pm$1.13 & 3.63 & 3.34$\pm$1.11 & \nodata & \nodata & \nodata & 0.076 \\
{\bf J0100+2802-SMG1 } & 01h00m13.9s & +28d00m25.8s & 9.94$\pm$1.47 & 6.77 & 10.00$\pm$2.08 & \nodata & \nodata & \nodata & 2.013 \\
{\bf J0100+2802-SMG2 } & 01h00m09.0s & +28d04m13.8s & 8.00$\pm$1.32 & 6.05 & 8.03$\pm$1.95 & \nodata & \nodata & \nodata & 2.050 \\
{\bf J0100+2802-SMG3 } & 01h00m14.2s & +28d02m53.8s & 5.48$\pm$1.12 & 4.88 & 5.18$\pm$1.50 & \nodata & \nodata & \nodata & 0.556 \\
{\bf J0100+2802-SMG4 } & 01h00m21.4s & +28d02m21.8s & 5.99$\pm$1.40 & 4.29 & 5.81$\pm$1.79 & \nodata & \nodata & \nodata & 2.116 \\
{\bf J0100+2802-SMG5 } & 01h00m17.5s & +28d03m17.8s & 4.11$\pm$1.16 & 3.53 & 3.37$\pm$1.14 & \nodata & \nodata & \nodata & 1.426 \\
\hline
{\bf J0148+0600-AGN} & 01h48m37.5s & +06d00m20.0s & 5.28$\pm$1.19 & 4.45 & 4.78$\pm$1.45 & \nodata & \nodata & \nodata & 0.000 \\
{\bf J0148+0600-SMG1 } & 01h48m37.8s & +05d58m28.0s & 6.83$\pm$1.49 & 4.57 & 6.76$\pm$2.05 & \nodata & \nodata & \nodata & 1.868 \\
{\bf J0148+0600-SMG2 } & 01h48m40.0s & +05d59m36.0s & 5.43$\pm$1.27 & 4.27 & 4.98$\pm$1.55 & \nodata & \nodata & \nodata & 0.950 \\
{\bf J0148+0600-SMG3 } & 01h48m39.4s & +06d00m44.0s & 5.18$\pm$1.22 & 4.23 & 4.66$\pm$1.45 & \nodata & \nodata & \nodata & 0.617 \\
{\bf J0148+0600-SMG4 } & 01h48m42.9s & +06d01m08.0s & 5.86$\pm$1.43 & 4.11 & 5.56$\pm$1.79 & \nodata & \nodata & \nodata & 1.561 \\
{\bf J0148+0600-SMG5 } & 01h48m30.8s & +06d00m12.0s & 5.23$\pm$1.42 & 3.67 & 4.72$\pm$1.61 & \nodata & \nodata & \nodata & 1.681 \\
{\bf J0148+0600-SMG6 } & 01h48m41.0s & +06d01m20.0s & 4.79$\pm$1.32 & 3.64 & 4.12$\pm$1.39 & \nodata & \nodata & \nodata & 1.326 \\
\hline
{\bf P036+03-AGN} & 02h26m02.8s & +03d02m59.4s & 5.47$\pm$1.01 & 5.42 & 5.31$\pm$1.48 & \nodata & \nodata & \nodata & 0.000 \\
{\bf P036+03-SMG1 } & 02h26m02.6s & +03d04m19.4s & 11.32$\pm$1.08 & 10.49 & 11.39$\pm$1.77 & \nodata & \nodata & \nodata & 1.335 \\
{\bf P036+03-SMG2 } & 02h25m58.0s & +03d03m47.4s & 6.33$\pm$1.06 & 6.00 & 6.33$\pm$1.68 & \nodata & \nodata & \nodata & 1.444 \\
{\bf P036+03-SMG3 } & 02h25m58.6s & +03d01m47.4s & 4.79$\pm$1.15 & 4.16 & 4.44$\pm$1.39 & \nodata & \nodata & \nodata & 1.607 \\
{\bf P036+03-SMG4 } & 02h26m10.0s & +03d02m43.4s & 5.20$\pm$1.31 & 3.96 & 4.96$\pm$1.62 & \nodata & \nodata & \nodata & 1.822 \\
{\bf P036+03-SMG5 } & 02h26m02.6s & +03d03m47.4s & 4.04$\pm$1.04 & 3.89 & 3.43$\pm$1.10 & 53.42$\pm$17.57 & 3.04 & 39.99$\pm$14.54 & 0.803 \\
{\bf P036+03-SMG6 } & 02h26m05.0s & +03d04m27.4s & 4.40$\pm$1.16 & 3.79 & 3.93$\pm$1.29 & \nodata & \nodata & \nodata & 1.561 \\
\hline
{\bf J0305-3150-AGN} & 03h05m18.2s & -31d50m55.8s & 8.43$\pm$1.08 & 7.78 & 8.50$\pm$1.77 & \nodata & \nodata & \nodata & 0.078 \\
{\bf J0305-3150-SMG1 } & 03h05m22.3s & -31d49m47.8s & 6.50$\pm$1.30 & 5.00 & 6.49$\pm$1.83 & \nodata & \nodata & \nodata & 1.579 \\
{\bf J0305-3150-SMG2 } & 03h05m12.5s & -31d51m23.8s & 5.27$\pm$1.11 & 4.76 & 5.00$\pm$1.47 & \nodata & \nodata & \nodata & 1.413 \\
{\bf J0305-3150-SMG3 } & 03h05m26.0s & -31d49m59.8s & 5.93$\pm$1.45 & 4.08 & 5.83$\pm$1.86 & \nodata & \nodata & \nodata & 2.244 \\
{\bf J0305-3150-SMG4 } & 03h05m07.8s & -31d50m51.8s & 4.41$\pm$1.23 & 3.58 & 3.86$\pm$1.31 & \nodata & \nodata & \nodata & 2.512 \\
\hline
{\bf P055-00-SMG1 } & 03h41m44.2s & -00d48m28.6s & 6.38$\pm$1.31 & 4.86 & 6.21$\pm$1.79 & \nodata & \nodata & \nodata & 0.657 \\
{\bf P055-00-SMG2 } & 03h41m40.8s & -00d47m12.6s & 4.84$\pm$1.25 & 3.88 & 4.21$\pm$1.35 & \nodata & \nodata & \nodata & 1.035 \\
{\bf P055-00-SMG3 } & 03h41m43.2s & -00d47m28.6s & 4.86$\pm$1.28 & 3.80 & 4.23$\pm$1.38 & \nodata & \nodata & \nodata & 0.806 \\
{\bf P055-00-SMG4 } & 03h41m38.1s & -00d50m04.6s & 5.39$\pm$1.49 & 3.61 & 4.97$\pm$1.69 & \nodata & \nodata & \nodata & 2.087 \\
{\bf P055-00-SMG5 } & 03h41m38.1s & -00d46m36.6s & 4.66$\pm$1.33 & 3.51 & 3.97$\pm$1.36 & \nodata & \nodata & \nodata & 1.853 \\
\hline
{\bf P056-16-SMG1 } & 03h46m58.0s & -16d29m44.8s & 5.85$\pm$1.05 & 5.55 & 5.81$\pm$1.51 & \nodata & \nodata & \nodata & 1.688 \\
{\bf P056-16-SMG2 } & 03h46m58.0s & -16d28m08.8s & 5.08$\pm$1.02 & 4.96 & 4.95$\pm$1.40 & \nodata & \nodata & \nodata & 1.336 \\
{\bf P056-16-SMG3 } & 03h46m48.3s & -16d27m04.8s & 4.16$\pm$1.09 & 3.81 & 3.84$\pm$1.27 & \nodata & \nodata & \nodata & 1.936 \\
\hline
{\bf P060+24-SMG1 } & 04h02m08.4s & +24d51m32.5s & 6.90$\pm$1.28 & 5.39 & 6.83$\pm$1.86 & \nodata & \nodata & \nodata & 1.329 \\
{\bf P060+24-SMG2 } & 04h02m07.2s & +24d50m08.5s & 6.78$\pm$1.46 & 4.65 & 6.69$\pm$1.97 & \nodata & \nodata & \nodata & 2.054 \\
{\bf P060+24-SMG3 } & 04h02m19.8s & +24d52m20.5s & 4.88$\pm$1.34 & 3.63 & 4.32$\pm$1.46 & \nodata & \nodata & \nodata & 1.803 \\
{\bf P060+24-SMG4 } & 04h02m12.2s & +24d52m48.5s & 4.41$\pm$1.23 & 3.57 & 3.68$\pm$1.25 & \nodata & \nodata & \nodata & 1.447 \\
\hline
{\bf P065-19-SMG1 } & 04h22m01.7s & -19d29m28.6s & 6.22$\pm$1.22 & 5.09 & 6.20$\pm$1.73 & \nodata & \nodata & \nodata & 2.001 \\
{\bf P065-19-SMG2 } & 04h21m56.9s & -19d26m28.6s & 5.13$\pm$1.04 & 4.95 & 4.95$\pm$1.45 & \nodata & \nodata & \nodata & 1.619 \\
{\bf P065-19-SMG3 } & 04h22m05.4s & -19d26m48.6s & 4.28$\pm$1.08 & 3.95 & 3.81$\pm$1.23 & \nodata & \nodata & \nodata & 1.079 \\
{\bf P065-19-SMG4 } & 04h22m09.3s & -19d26m20.6s & 4.98$\pm$1.32 & 3.78 & 4.74$\pm$1.60 & \nodata & \nodata & \nodata & 2.160 \\
\hline
{\bf P089-15-AGN} & 05h59m46.4s & -15d35m00.1s & 3.56$\pm$1.17 & 3.05 & 2.60$\pm$0.96 & \nodata & \nodata & \nodata & 0.000 \\
{\bf P089-15-SMG1 } & 05h59m42.8s & -15d34m56.1s & 6.28$\pm$1.17 & 5.36 & 6.13$\pm$1.65 & 100.56$\pm$31.85 & 3.16 & 76.23$\pm$27.15 & 0.902 \\
{\bf P089-15-SMG2 } & 05h59m53.3s & -15d34m20.1s & 5.64$\pm$1.46 & 3.86 & 5.37$\pm$1.75 & 113.21$\pm$36.11 & 3.13 & 93.10$\pm$33.58 & 1.854 \\
\hline
{\bf J0810+5105-SMG1 } & 08h10m57.6s & +51d04m32.1s & 4.85$\pm$1.22 & 3.98 & 4.34$\pm$1.38 & \nodata & \nodata & \nodata & 1.416 \\
{\bf J0810+5105-SMG2 } & 08h10m52.1s & +51d04m40.1s & 4.65$\pm$1.26 & 3.69 & 4.06$\pm$1.35 & \nodata & \nodata & \nodata & 1.132 \\
{\bf J0810+5105-SMG3 } & 08h11m04.4s & +51d04m52.1s & 4.70$\pm$1.28 & 3.69 & 4.13$\pm$1.37 & \nodata & \nodata & \nodata & 2.670 \\
{\bf J0810+5105-SMG4 } & 08h10m47.5s & +51d06m36.1s & 4.76$\pm$1.34 & 3.55 & 4.22$\pm$1.44 & \nodata & \nodata & \nodata & 1.938 \\
\hline
{\bf J0828+2633-SMG1 } & 08h28m09.5s & +26d34m39.5s & 7.12$\pm$1.31 & 5.45 & 7.10$\pm$1.88 & \nodata & \nodata & \nodata & 1.215 \\
{\bf J0828+2633-SMG2 } & 08h28m15.1s & +26d31m55.5s & 5.68$\pm$1.45 & 3.92 & 5.41$\pm$1.75 & \nodata & \nodata & \nodata & 2.049 \\
{\bf J0828+2633-SMG3 } & 08h28m15.4s & +26d33m03.5s & 4.41$\pm$1.23 & 3.58 & 3.71$\pm$1.24 & \nodata & \nodata & \nodata & 1.012 \\
\hline
{\bf J0835+3217-SMG1 } & 08h35m33.3s & +32d19m08.6s & 5.89$\pm$1.42 & 4.15 & 5.65$\pm$1.75 & \nodata & \nodata & \nodata & 2.278 \\
{\bf J0835+3217-SMG2 } & 08h35m34.9s & +32d18m00.6s & 5.79$\pm$1.49 & 3.89 & 5.52$\pm$1.78 & \nodata & \nodata & \nodata & 2.291 \\
{\bf J0835+3217-SMG3 } & 08h35m32.1s & +32d17m56.6s & 5.08$\pm$1.35 & 3.78 & 4.59$\pm$1.49 & 54.28$\pm$17.32 & 3.13 & 43.16$\pm$15.51 & 1.579 \\
\hline
{\bf J0839+0015-SMG1 } & 08h39m57.7s & +00d16m50.2s & 4.75$\pm$1.28 & 3.72 & 4.18$\pm$1.37 & \nodata & \nodata & \nodata & 1.110 \\
\hline
{\bf J0842+1218-SMG1 } & 08h42m27.4s & +12d19m34.5s & 8.49$\pm$1.36 & 6.24 & 8.53$\pm$2.05 & \nodata & \nodata & \nodata & 0.875 \\
{\bf J0842+1218-SMG2 } & 08h42m25.8s & +12d18m50.5s & 6.15$\pm$1.31 & 4.68 & 5.91$\pm$1.74 & 51.62$\pm$13.79 & 3.74 & 44.90$\pm$14.33 & 0.887 \\
\hline
{\bf J0850+3246-SMG1 } & 08h50m41.9s & +32d45m11.9s & 6.08$\pm$1.39 & 4.38 & 5.86$\pm$1.82 & \nodata & \nodata & \nodata & 2.253 \\
{\bf J0850+3246-SMG2 } & 08h50m54.3s & +32d46m51.9s & 4.79$\pm$1.35 & 3.54 & 4.12$\pm$1.41 & \nodata & \nodata & \nodata & 1.508 \\
\hline
{\bf P135+16-AGN} & 09h01m32.6s & +16d15m06.8s & 5.16$\pm$1.34 & 3.86 & 4.49$\pm$1.44 & \nodata & \nodata & \nodata & 0.000 \\
\hline
{\bf P159-02-SMG1 } & 10h36m48.3s & -02d31m45.8s & 6.61$\pm$1.38 & 4.80 & 6.47$\pm$1.89 & \nodata & \nodata & \nodata & 1.705 \\
{\bf P159-02-SMG2 } & 10h36m58.7s & -02d31m49.8s & 5.51$\pm$1.49 & 3.69 & 5.05$\pm$1.70 & \nodata & \nodata & \nodata & 1.388 \\
\hline
{\bf J1048-0109-AGN} & 10h48m19.0s & -01d09m40.2s & 4.56$\pm$1.17 & 3.89 & 3.99$\pm$1.26 & \nodata & \nodata & \nodata & 0.000 \\
{\bf J1048-0109-SMG1 } & 10h48m20.4s & -01d11m12.2s & 7.74$\pm$1.37 & 5.66 & 7.79$\pm$1.91 & \nodata & \nodata & \nodata & 1.569 \\
{\bf J1048-0109-SMG2 } & 10h48m23.0s & -01d09m00.2s & 5.61$\pm$1.33 & 4.23 & 5.37$\pm$1.63 & \nodata & \nodata & \nodata & 1.202 \\
{\bf J1048-0109-SMG3 } & 10h48m12.1s & -01d08m20.2s & 4.97$\pm$1.39 & 3.59 & 4.54$\pm$1.51 & \nodata & \nodata & \nodata & 2.187 \\
{\bf J1048-0109-SMG4 } & 10h48m22.8s & -01d10m20.2s & 4.59$\pm$1.30 & 3.53 & 4.03$\pm$1.35 & \nodata & \nodata & \nodata & 1.147 \\
\hline
{\bf P167-13-SMG1 } & 11h10m30.7s & -13d31m05.5s & 5.40$\pm$1.38 & 3.92 & 5.03$\pm$1.62 & \nodata & \nodata & \nodata & 1.567 \\
{\bf P167-13-SMG2 } & 11h10m29.6s & -13d29m09.5s & 4.31$\pm$1.21 & 3.56 & 3.58$\pm$1.20 & \nodata & \nodata & \nodata & 1.250 \\
\hline
{\bf J1143+3808-SMG1 } & 11h43m46.0s & +38d07m48.7s & 6.19$\pm$1.49 & 4.15 & 5.85$\pm$1.83 & \nodata & \nodata & \nodata & 2.060 \\
{\bf J1143+3808-SMG2 } & 11h43m37.6s & +38d09m36.7s & 5.86$\pm$1.43 & 4.10 & 5.44$\pm$1.70 & \nodata & \nodata & \nodata & 1.146 \\
\hline
{\bf J1148+0702-SMG1 } & 11h48m00.6s & +07d02m44.3s & 10.02$\pm$1.33 & 7.54 & 10.07$\pm$1.99 & \nodata & \nodata & \nodata & 0.901 \\
{\bf J1148+0702-SMG2 } & 11h48m05.4s & +07d02m48.3s & 4.87$\pm$1.36 & 3.58 & 4.22$\pm$1.39 & \nodata & \nodata & \nodata & 0.856 \\
\hline
{\bf J1152+0055-SMG1 } & 11h52m12.9s & +00d55m48.6s & 7.68$\pm$1.23 & 6.23 & 7.62$\pm$1.84 & \nodata & \nodata & \nodata & 2.077 \\
{\bf J1152+0055-SMG2 } & 11h52m23.4s & +00d56m28.6s & 4.16$\pm$1.13 & 3.69 & 3.48$\pm$1.17 & \nodata & \nodata & \nodata & 1.018 \\
{\bf J1152+0055-SMG3 } & 11h52m19.1s & +00d57m00.6s & 4.19$\pm$1.15 & 3.65 & 3.52$\pm$1.20 & \nodata & \nodata & \nodata & 1.498 \\
{\bf J1152+0055-SMG4 } & 11h52m21.0s & +00d56m28.6s & 3.98$\pm$1.09 & 3.65 & 3.24$\pm$1.09 & \nodata & \nodata & \nodata & 0.869 \\
{\bf J1152+0055-SMG5 } & 11h52m28.2s & +00d56m48.6s & 5.32$\pm$1.50 & 3.56 & 5.01$\pm$1.76 & \nodata & \nodata & \nodata & 2.108 \\
\hline
{\bf J1205+0505-SMG1 } & 12h05m06.4s & +00d01m03.9s & 7.49$\pm$1.31 & 5.72 & 7.49$\pm$1.85 & \nodata & \nodata & \nodata & 1.569 \\
{\bf J1205+0505-SMG2 } & 12h05m08.0s & +00d00m03.9s & 5.12$\pm$1.32 & 3.89 & 4.69$\pm$1.49 & \nodata & \nodata & \nodata & 0.907 \\
\hline
{\bf J1207+0630-SMG1 } & 12h07m36.3s & +06d31m46.0s & 6.03$\pm$1.30 & 4.62 & 5.85$\pm$1.72 & 31.86$\pm$10.23 & 3.11 & 25.61$\pm$9.28 & 1.622 \\
{\bf J1207+0630-SMG2 } & 12h07m37.9s & +06d30m50.0s & 4.47$\pm$1.24 & 3.59 & 3.73$\pm$1.22 & \nodata & \nodata & \nodata & 0.680 \\
\hline
{\bf P183+05-AGN} & 12h12m27.2s & +05d05m29.4s & 9.03$\pm$1.30 & 6.93 & 9.13$\pm$2.03 & \nodata & \nodata & \nodata & 0.094 \\
{\bf P183+05-SMG1 } & 12h12m23.4s & +05d05m33.4s & 4.77$\pm$1.34 & 3.55 & 4.06$\pm$1.37 & \nodata & \nodata & \nodata & 0.870 \\
{\bf P183+05-SMG2 } & 12h12m29.6s & +05d06m13.4s & 4.93$\pm$1.42 & 3.48 & 4.30$\pm$1.48 & 33.47$\pm$8.93 & 3.75 & 28.82$\pm$9.26 & 0.945 \\
\hline
{\bf P183-12-AGN} & 12h13m11.7s & -12d46m03.5s & 4.08$\pm$1.14 & 3.59 & 3.25$\pm$1.09 & \nodata & \nodata & \nodata & 0.000 \\
{\bf P183-12-SMG1 } & 12h13m13.9s & -12d46m39.5s & 9.48$\pm$1.21 & 7.83 & 9.48$\pm$1.86 & \nodata & \nodata & \nodata & 0.812 \\
{\bf P183-12-SMG2 } & 12h13m14.8s & -12d45m11.5s & 4.53$\pm$1.25 & 3.62 & 3.89$\pm$1.32 & \nodata & \nodata & \nodata & 1.147 \\
\hline
{\bf P184+01-AGN} & 12h17m21.8s & +01d31m30.4s & 3.67$\pm$1.12 & 3.26 & 2.76$\pm$0.97 & \nodata & \nodata & \nodata & 0.240 \\
{\bf P184+01-SMG1 } & 12h17m26.6s & +01d31m14.4s & 5.85$\pm$1.29 & 4.54 & 5.61$\pm$1.69 & 43.03$\pm$11.41 & 3.77 & 39.76$\pm$12.68 & 1.413 \\
{\bf P184+01-SMG2 } & 12h17m18.6s & +01d30m50.4s & 4.96$\pm$1.17 & 4.23 & 4.48$\pm$1.39 & \nodata & \nodata & \nodata & 1.094 \\
{\bf P184+01-SMG3 } & 12h17m20.0s & +01d33m10.4s & 4.55$\pm$1.24 & 3.67 & 3.91$\pm$1.31 & \nodata & \nodata & \nodata & 1.504 \\
\hline
{\bf P187+04-AGN} & 12h29m12.3s & +04d19m23.6s & 3.81$\pm$1.25 & 3.04 & 2.79$\pm$1.03 & \nodata & \nodata & \nodata & 0.211 \\
{\bf P187+04-SMG1 } & 12h29m09.7s & +04d20m47.6s & 6.88$\pm$1.35 & 5.09 & 6.79$\pm$1.92 & \nodata & \nodata & \nodata & 1.592 \\
{\bf P187+04-SMG2 } & 12h29m13.1s & +04d20m03.6s & 5.35$\pm$1.24 & 4.30 & 4.84$\pm$1.46 & \nodata & \nodata & \nodata & 0.600 \\
\hline
{\bf J1243+2529-SMG1 } & 12h43m43.4s & +25d28m15.8s & 7.94$\pm$1.45 & 5.46 & 8.01$\pm$2.17 & \nodata & \nodata & \nodata & 1.314 \\
{\bf J1243+2529-SMG2 } & 12h43m42.8s & +25d30m19.8s & 5.55$\pm$1.40 & 3.95 & 5.06$\pm$1.61 & \nodata & \nodata & \nodata & 1.067 \\
\hline
{\bf J1257+6349-SMG1 } & 12h57m59.3s & +63d50m29.1s & 5.11$\pm$1.10 & 4.65 & 4.84$\pm$1.39 & \nodata & \nodata & \nodata & 0.978 \\
{\bf J1257+6349-SMG2 } & 12h57m46.6s & +63d49m49.1s & 5.29$\pm$1.20 & 4.42 & 5.07$\pm$1.50 & 35.14$\pm$9.23 & 3.81 & 32.30$\pm$10.30 & 2.728 \\
\hline
{\bf P210+27-SMG1 } & 14h01m53.0s & +27d49m10.9s & 9.33$\pm$1.30 & 7.15 & 9.45$\pm$2.13 & \nodata & \nodata & \nodata & 1.487 \\
{\bf P210+27-SMG2 } & 14h01m46.1s & +27d49m54.9s & 5.02$\pm$1.20 & 4.19 & 4.46$\pm$1.41 & \nodata & \nodata & \nodata & 0.449 \\
{\bf P210+27-SMG3 } & 14h01m45.5s & +27d48m46.9s & 4.75$\pm$1.18 & 4.03 & 4.09$\pm$1.31 & 38.46$\pm$10.43 & 3.69 & 32.77$\pm$10.57 & 0.919 \\
{\bf P210+27-SMG4 } & 14h01m49.4s & +27d49m38.9s & 4.47$\pm$1.19 & 3.74 & 3.70$\pm$1.22 & \nodata & \nodata & \nodata & 0.532 \\
{\bf P210+27-SMG5 } & 14h01m39.8s & +27d50m18.9s & 5.29$\pm$1.49 & 3.54 & 4.84$\pm$1.71 & \nodata & \nodata & \nodata & 2.022 \\
{\bf P210+27-SMG6 } & 14h01m40.1s & +27d48m54.9s & 4.99$\pm$1.42 & 3.51 & 4.43$\pm$1.56 & \nodata & \nodata & \nodata & 1.928 \\
{\bf P210+27-SMG7 } & 14h01m42.8s & +27d49m38.9s & 4.47$\pm$1.28 & 3.50 & 3.70$\pm$1.28 & \nodata & \nodata & \nodata & 1.133 \\
\hline
{\bf P210+40-SMG1 } & 14h03m00.6s & +40d23m59.1s & 5.51$\pm$1.27 & 4.34 & 5.18$\pm$1.57 & \nodata & \nodata & \nodata & 1.490 \\
{\bf P210+40-SMG2 } & 14h02m55.3s & +40d24m15.1s & 4.69$\pm$1.24 & 3.77 & 4.04$\pm$1.31 & \nodata & \nodata & \nodata & 0.266 \\
\hline
\hline
{\bf P210-12-AGN} & 14h03m29.8s & -12d00m34.1s & 3.56$\pm$1.16 & 3.07 & 2.61$\pm$0.96 & \nodata & \nodata & \nodata & 0.136 \\
{\bf P210-12-SMG1 } & 14h03m31.4s & -12d01m06.1s & 5.71$\pm$1.21 & 4.70 & 5.45$\pm$1.58 & \nodata & \nodata & \nodata & 0.763 \\
{\bf P210-12-SMG2 } & 14h03m28.2s & -11d58m50.1s & 5.41$\pm$1.40 & 3.85 & 5.06$\pm$1.64 & \nodata & \nodata & \nodata & 1.755 \\
\hline
{\bf P215-16-AGN} & 14h20m36.2s & -16d02m30.2s & 16.85$\pm$1.10 & 15.36 & 16.96$\pm$1.90 & 26.93$\pm$7.78 & 3.46 & 21.66$\pm$7.26 & 0.000 \\
{\bf P215-16-SMG1 } & 14h20m35.4s & -16d03m34.2s & 6.08$\pm$1.20 & 5.07 & 5.97$\pm$1.73 & \nodata & \nodata & \nodata & 1.087 \\
{\bf P215-16-SMG2 } & 14h20m37.1s & -16d01m58.2s & 4.97$\pm$1.14 & 4.36 & 4.54$\pm$1.40 & 27.51$\pm$7.92 & 3.47 & 22.46$\pm$7.53 & 0.572 \\
{\bf P215-16-SMG3 } & 14h20m35.7s & -16d00m50.2s & 5.13$\pm$1.30 & 3.95 & 4.76$\pm$1.56 & \nodata & \nodata & \nodata & 1.672 \\
{\bf P215-16-SMG4 } & 14h20m34.0s & -16d02m02.2s & 3.97$\pm$1.10 & 3.60 & 3.17$\pm$1.06 & \nodata & \nodata & \nodata & 0.725 \\
{\bf P215-16-SMG5 } & 14h20m39.6s & -16d02m02.2s & 4.19$\pm$1.19 & 3.54 & 3.47$\pm$1.19 & 27.06$\pm$8.28 & 3.27 & 21.87$\pm$7.65 & 0.954 \\
\hline
{\bf P215+26-SMG1 } & 14h21m42.0s & +26d33m09.0s & 6.83$\pm$1.22 & 5.59 & 6.75$\pm$1.79 & \nodata & \nodata & \nodata & 1.236 \\
{\bf P215+26-SMG2 } & 14h21m42.3s & +26d33m45.0s & 5.79$\pm$1.28 & 4.51 & 5.52$\pm$1.65 & \nodata & \nodata & \nodata & 1.814 \\
{\bf P215+26-SMG3 } & 14h21m43.2s & +26d31m33.0s & 4.56$\pm$1.21 & 3.76 & 3.89$\pm$1.26 & 33.35$\pm$10.85 & 3.07 & 24.92$\pm$9.01 & 0.400 \\
{\bf P215+26-SMG4 } & 14h21m50.1s & +26d33m01.0s & 5.11$\pm$1.43 & 3.57 & 4.62$\pm$1.57 & \nodata & \nodata & \nodata & 2.019 \\
{\bf P215+26-SMG5 } & 14h21m37.9s & +26d32m41.0s & 4.45$\pm$1.26 & 3.53 & 3.74$\pm$1.26 & \nodata & \nodata & \nodata & 1.529 \\
\hline
{\bf P217-16-SMG1 } & 14h28m21.3s & -16d01m15.3s & 5.81$\pm$1.38 & 4.22 & 5.54$\pm$1.73 & \nodata & \nodata & \nodata & 1.467 \\
{\bf P217-16-SMG2 } & 14h28m20.2s & -16d02m15.3s & 4.52$\pm$1.22 & 3.69 & 3.79$\pm$1.24 & \nodata & \nodata & \nodata & 0.543 \\
\hline
{\bf P217-07-AGN} & 14h31m40.7s & -07d24m47.5s & 6.03$\pm$1.17 & 5.16 & 5.86$\pm$1.61 & \nodata & \nodata & \nodata & 0.095 \\
{\bf P217-07-SMG1 } & 14h31m38.0s & -07d24m03.5s & 4.97$\pm$1.22 & 4.08 & 4.53$\pm$1.40 & \nodata & \nodata & \nodata & 0.900 \\
\hline
{\bf P231-20-AGN} & 15h26m37.8s & -20d50m00.7s & 7.99$\pm$1.22 & 6.53 & 8.06$\pm$1.99 & 80.31$\pm$19.97 & 4.02 & 71.00$\pm$21.54 & 0.000 \\
{\bf P231-20-SMG1 } & 15h26m39.8s & -20d51m12.7s & 6.01$\pm$1.35 & 4.46 & 5.77$\pm$1.77 & \nodata & \nodata & \nodata & 1.300 \\
{\bf P231-20-SMG2 } & 15h26m38.4s & -20d50m44.7s & 5.40$\pm$1.26 & 4.27 & 4.98$\pm$1.56 & \nodata & \nodata & \nodata & 0.747 \\
{\bf P231-20-SMG3 } & 15h26m38.9s & -20d51m36.7s & 5.57$\pm$1.39 & 4.00 & 5.22$\pm$1.70 & \nodata & \nodata & \nodata & 1.625 \\
\hline
{\bf P239-07-SMG1 } & 15h58m56.0s & -07d23m57.6s & 4.62$\pm$1.31 & 3.53 & 4.02$\pm$1.37 & \nodata & \nodata & \nodata & 1.293 \\
\hline
{\bf J1609+3041-AGN} & 16h09m36.9s & +30d41m47.6s & 4.09$\pm$1.17 & 3.50 & 3.27$\pm$1.11 & \nodata & \nodata & \nodata & 0.078 \\
{\bf J1609+3041-SMG1 } & 16h09m44.7s & +30d41m51.6s & 7.19$\pm$1.44 & 5.01 & 7.16$\pm$2.00 & 25.56$\pm$8.27 & 3.09 & 20.53$\pm$7.45 & 1.862 \\
{\bf J1609+3041-SMG2 } & 16h09m32.3s & +30d43m43.6s & 5.40$\pm$1.35 & 3.99 & 5.02$\pm$1.62 & \nodata & \nodata & \nodata & 2.297 \\
{\bf J1609+3041-SMG3 } & 16h09m42.5s & +30d43m03.6s & 4.90$\pm$1.28 & 3.83 & 4.37$\pm$1.44 & \nodata & \nodata & \nodata & 1.828 \\
{\bf J1609+3041-SMG4 } & 16h09m31.4s & +30d42m43.6s & 4.63$\pm$1.28 & 3.62 & 4.01$\pm$1.36 & \nodata & \nodata & \nodata & 1.744 \\
\hline
{\bf P247+24-AGN} & 16h29m11.8s & +24d07m35.6s & 6.76$\pm$1.13 & 5.98 & 6.77$\pm$1.72 & \nodata & \nodata & \nodata & 0.161 \\
{\bf P247+24-SMG1 } & 16h29m03.7s & +24d08m39.6s & 6.65$\pm$1.32 & 5.05 & 6.65$\pm$1.83 & \nodata & \nodata & \nodata & 2.146 \\
{\bf P247+24-SMG2 } & 16h29m05.1s & +24d06m07.6s & 6.65$\pm$1.39 & 4.80 & 6.66$\pm$1.89 & \nodata & \nodata & \nodata & 2.169 \\
\hline
\hline
{\bf P308-21-AGN} & 20h32m09.7s & -21d14m02.3s & 4.23$\pm$1.09 & 3.86 & 3.59$\pm$1.17 & \nodata & \nodata & \nodata & 0.072 \\
{\bf P308-21-SMG1 } & 20h32m15.9s & -21d13m02.3s & 9.92$\pm$1.37 & 7.27 & 10.00$\pm$2.00 & \nodata & \nodata & \nodata & 1.804 \\
{\bf P308-21-SMG2 } & 20h32m10.8s & -21d12m54.3s & 6.13$\pm$1.22 & 5.02 & 6.02$\pm$1.71 & 60.37$\pm$18.27 & 3.30 & 49.54$\pm$17.18 & 1.153 \\
{\bf P308-21-SMG3 } & 20h32m05.1s & -21d15m10.3s & 5.03$\pm$1.19 & 4.23 & 4.65$\pm$1.44 & \nodata & \nodata & \nodata & 1.662 \\
{\bf P308-21-SMG4 } & 20h32m06.8s & -21d14m10.3s & 4.30$\pm$1.11 & 3.88 & 3.70$\pm$1.20 & \nodata & \nodata & \nodata & 0.798 \\
{\bf P308-21-SMG5 } & 20h32m08.8s & -21d16m02.3s & 4.72$\pm$1.30 & 3.64 & 4.27$\pm$1.45 & 62.30$\pm$20.10 & 3.10 & 52.20$\pm$19.02 & 2.020 \\
{\bf P308-21-SMG6 } & 20h32m08.8s & -21d13m10.3s & 4.32$\pm$1.19 & 3.64 & 3.73$\pm$1.26 & \nodata & \nodata & \nodata & 0.913 \\
{\bf P308-21-SMG7 } & 20h32m13.7s & -21d15m06.3s & 4.50$\pm$1.24 & 3.63 & 3.97$\pm$1.35 & \nodata & \nodata & \nodata & 1.415 \\
{\bf P308-21-SMG8 } & 20h32m13.7s & -21d14m26.3s & 4.17$\pm$1.18 & 3.55 & 3.51$\pm$1.20 & \nodata & \nodata & \nodata & 1.012 \\
\hline
{\bf J2100-1715-SMG1 } & 21h01m00.0s & -17d14m02.5s & 5.21$\pm$1.20 & 4.34 & 5.05$\pm$1.55 & \nodata & \nodata & \nodata & 1.739 \\
{\bf J2100-1715-SMG2 } & 21h00m58.1s & -17d16m58.5s & 4.68$\pm$1.12 & 4.18 & 4.40$\pm$1.39 & \nodata & \nodata & \nodata & 1.719 \\
{\bf J2100-1715-SMG3 } & 21h00m53.6s & -17d15m14.5s & 3.73$\pm$0.95 & 3.91 & 3.18$\pm$1.04 & \nodata & \nodata & \nodata & 0.507 \\
{\bf J2100-1715-SMG4 } & 21h00m58.4s & -17d13m42.5s & 4.76$\pm$1.22 & 3.89 & 4.52$\pm$1.49 & \nodata & \nodata & \nodata & 1.807 \\
\hline
{\bf P323+12-SMG1 } & 21h32m35.2s & +12d16m27.1s & 5.78$\pm$1.24 & 4.68 & 5.66$\pm$1.65 & \nodata & \nodata & \nodata & 1.492 \\
{\bf P323+12-SMG2 } & 21h32m25.7s & +12d18m43.1s & 4.63$\pm$1.24 & 3.72 & 4.24$\pm$1.42 & \nodata & \nodata & \nodata & 2.261 \\
{\bf P323+12-SMG3 } & 21h32m28.7s & +12d16m39.1s & 4.46$\pm$1.22 & 3.65 & 4.03$\pm$1.37 & \nodata & \nodata & \nodata & 1.862 \\
\hline
{\bf P333+26-AGN} & 22h15m57.6s & +26d06m33.3s & 3.83$\pm$1.04 & 3.69 & 3.07$\pm$1.03 & \nodata & \nodata & \nodata & 0.067 \\
{\bf P333+26-SMG1 } & 22h16m05.9s & +26d07m21.3s & 8.11$\pm$1.27 & 6.39 & 8.12$\pm$1.89 & \nodata & \nodata & \nodata & 2.252 \\
{\bf P333+26-SMG2 } & 22h16m04.4s & +26d07m13.3s & 5.70$\pm$1.20 & 4.74 & 5.54$\pm$1.68 & \nodata & \nodata & \nodata & 1.858 \\
{\bf P333+26-SMG3 } & 22h16m03.8s & +26d07m53.3s & 4.79$\pm$1.19 & 4.01 & 4.37$\pm$1.43 & 95.40$\pm$22.93 & 4.16 & 92.95$\pm$27.69 & 2.095 \\
{\bf P333+26-SMG4 } & 22h15m50.5s & +26d05m33.3s & 4.65$\pm$1.22 & 3.82 & 4.18$\pm$1.41 & \nodata & \nodata & \nodata & 2.011 \\
{\bf P333+26-SMG5 } & 22h15m54.3s & +26d07m45.3s & 4.22$\pm$1.13 & 3.73 & 3.60$\pm$1.22 & \nodata & \nodata & \nodata & 1.507 \\
\hline
{\bf P338+29-SMG1 } & 22h32m58.5s & +29d29m44.1s & 6.50$\pm$1.45 & 4.48 & 6.31$\pm$1.92 & \nodata & \nodata & \nodata & 1.162 \\
{\bf P338+29-SMG2 } & 22h32m54.5s & +29d32m08.1s & 6.16$\pm$1.43 & 4.30 & 5.88$\pm$1.83 & \nodata & \nodata & \nodata & 1.607 \\
{\bf P338+29-SMG3 } & 22h32m56.4s & +29d31m00.1s & 4.70$\pm$1.32 & 3.57 & 3.94$\pm$1.36 & \nodata & \nodata & \nodata & 0.558 \\
\hline
{\bf P340-18-SMG1 } & 22h40m46.9s & -18d39m43.8s & 4.91$\pm$1.03 & 4.79 & 4.64$\pm$1.36 & \nodata & \nodata & \nodata & 0.774 \\
{\bf P340-18-SMG2 } & 22h40m48.6s & -18d39m11.8s & 4.10$\pm$1.03 & 3.97 & 3.57$\pm$1.14 & \nodata & \nodata & \nodata & 0.639 \\
{\bf P340-18-SMG3 } & 22h40m49.1s & -18d37m39.8s & 4.93$\pm$1.31 & 3.76 & 4.66$\pm$1.56 & \nodata & \nodata & \nodata & 2.077 \\
{\bf P340-18-SMG4 } & 22h40m50.0s & -18d40m23.8s & 3.86$\pm$1.05 & 3.66 & 3.24$\pm$1.08 & \nodata & \nodata & \nodata & 0.667 \\
\hline
{\bf J2348-3054-AGN} & 23h48m34.3s & -30d54m10.1s & 5.88$\pm$1.06 & 5.52 & 5.81$\pm$1.51 & \nodata & \nodata & \nodata & 0.000 \\
{\bf J2348-3054-SMG1 } & 23h48m40.8s & -30d53m46.1s & 5.02$\pm$1.27 & 3.96 & 4.78$\pm$1.50 & \nodata & \nodata & \nodata & 1.680 \\
\hline
{\bf P359-06-SMG1 } & 23h56m30.7s & -06d23m35.2s & 4.50$\pm$1.14 & 3.96 & 3.95$\pm$1.28 & \nodata & \nodata & \nodata & 0.900 \\
{\bf P359-06-SMG2 } & 23h56m28.3s & -06d24m07.2s & 5.08$\pm$1.31 & 3.88 & 4.74$\pm$1.57 & \nodata & \nodata & \nodata & 1.706 \\
{\bf P359-06-SMG3 } & 23h56m27.5s & -06d22m31.2s & 4.35$\pm$1.16 & 3.76 & 3.73$\pm$1.24 & \nodata & \nodata & \nodata & 1.548 \\
{\bf P359-06-SMG4 } & 23h56m37.4s & -06d21m47.2s & 4.95$\pm$1.33 & 3.73 & 4.56$\pm$1.55 & \nodata & \nodata & \nodata & 1.566 \\
\hline
\enddata
\end{deluxetable*}
\end{longrotatetable}


\bibliography{sample631}{}
\bibliographystyle{aasjournal}



\end{document}